\documentclass[acmtog]{acmart}
\acmSubmissionID{papers 520}
\acmJournal{TOG}

\usepackage{booktabs} % For formal tables

\usepackage{microtype} % more sophisticated kerning, targeting uniform gray / straight margins
\usepackage{breakcites}
\usepackage{ellipsis} % fix spacing around ...
\usepackage[mathletters]{ucs}
\usepackage[utf8x]{inputenc}
\usepackage{mathtools}
\usepackage{amsbsy}
\usepackage{algorithm}
\usepackage{algpseudocode}
\usepackage{upgreek}
\usepackage{dblfloatfix}

\usepackage{tabularx}
\usepackage{booktabs}

\usepackage{amsmath}
\usepackage{amsfonts}
\usepackage{amssymb}

\definecolor{lightbluishgrey}{rgb}{0.78,0.86,0.93}
\newcommand{\change}[1]{#1}

\usepackage{layouts}

\usepackage{wrapfig}
\newcommand{\reffig}[1] {Fig.~\ref{fig:#1}}

\makeatletter
\def\reffig{\@ifnextchar[{\@myreffigloc}{\@myreffignoloc}}
\def\@myreffigloc[#1]#2{Figure~\ref{fig:#2}, \emph{#1}}
\def\@myreffignoloc#1{Figure~\ref{fig:#1}}
\makeatother
\newtheorem*{claim}{Claim}
\newtheorem*{unlemma}{Lemma}

\usepackage{xfrac}

\usepackage{amsmath}
\usepackage{amssymb}    % need for things like varnothing
\usepackage{cancel}
\usepackage[T1]{fontenc}
\newcommand{\mlp}{\boldsymbol{f}}
\newcommand{\proj}{\pi}
\newcommand{\codeword}{\boldsymbol{z}}
\newcommand{\centroid}{\boldsymbol{c}}
\newcommand{\pred}{{P}}
\newcommand{\rest}{{R}}
\newcommand{\tri}{{\boldsymbol{t}}}
\newcommand{\vertices}{\textbf{V}}
\newcommand{\triangles}{\textbf{T}}
\newcommand{\fram}{\mathcal{B}}
\newcommand{\source}{\mathcal{S}}
\newcommand{\jloss}{l_\text{jacobian}}
\newcommand{\vloss}{l_\text{vertex}}
\newcommand{\totloss}{l_{\text{total}}}
\newcommand{\jac}{J}
\newcommand{\tangent}{{T}}
\newcommand{\predmap}{\Phi}
\newcommand{\gtmap}{\Psi}

\newcommand{\vc}[1]{\mathbf{#1}}

\newcommand{\norm}[1]{\left\| #1\right\|}
\newcommand{\abs}[1]{\left\vert#1\right\vert}

\newcommand{\set}[1]{\left\{#1\right\}}
\newcommand{\parr}[1]{\left (#1\right )}
\newcommand{\brac}[1]{\left [#1\right ]}

\renewcommand{\t}{\vc{t}}
\usepackage{siunitx}
\usepackage{graphicx}

\usepackage{wrapfig}

\makeatletter

\usepackage{etoolbox}
\usepackage{array}

\newcommand{\figs}{}
\def\figs/{figs/}
\newcommand{\Real}{\mathbb R}

\let\min\relax
\DeclareMathOperator*{\min}{\text{min }}

\usepackage[verbose]{newunicodechar}
\usepackage{empheq}

\usepackage{xcolor}

\setcopyright{acmcopyright}\acmJournal{TOG}
\acmYear{2022}\acmVolume{41}\acmNumber{4}\acmArticle{109}\acmMonth{7} \acmDOI{10.1145/3528223.3530141}

\begin{document}

\title{Neural Jacobian Fields: Learning Intrinsic Mappings of Arbitrary Meshes
}

\author{Noam Aigerman}

\affiliation{%
   \institution{Adobe Research}
   \country{USA}
}
 \email{aigerman@adobe.com}
\author{Kunal Gupta}
\authornote{Participated as part of an internship at Adobe.}
\affiliation{%
   \institution{University of California San Diego}
   \country{USA}
}
\email{k5gupta@eng.ucsd.edu}

\author{Vladimir G. Kim}
 \email{vokim@adobe.com}
\affiliation{%
   \institution{Adobe Research}
   \country{USA}
}

\author{Siddhartha Chaudhuri}
 \email{sidch@adobe.com}
\affiliation{%
   \institution{Adobe Research, IIT Bombay}
   \country{India}
}

\author{Jun Saito}
 \email{jsaito@adobe.com}
\affiliation{%
   \institution{Adobe Research}
   \country{USA}
}
\author{Thibault Groueix}
 \email{groueix@adobe.com}
\affiliation{%
   \institution{Adobe Research}
   \country{USA}
}
\begin{abstract}
  This paper introduces a framework designed to accurately predict piecewise linear mappings of arbitrary meshes via a neural network, enabling training and evaluating over heterogeneous collections of meshes that do not share a triangulation, as well as producing highly detail-preserving maps whose accuracy exceeds current state of the art.  The framework is based on reducing the neural aspect to a prediction of a matrix for a  single given point, conditioned on a global shape descriptor. The field of matrices is then projected onto the tangent bundle of the given mesh, and used as candidate jacobians for the predicted map. The map is computed by a standard Poisson solve, implemented as a differentiable layer with cached pre-factorization for efficient training. This construction is agnostic to the triangulation of the input, thereby enabling applications on datasets with varying triangulations. At the same time,  by operating in the intrinsic gradient domain of each individual mesh, it allows the framework to predict highly-accurate mappings.  We validate these properties by conducting experiments over a broad range of scenarios, from semantic ones such as morphing, registration, and deformation transfer, to optimization-based ones, such as emulating elastic deformations and contact correction, as well as being the first work, to our knowledge, to tackle the task of learning to compute UV parameterizations of arbitrary meshes. The results exhibit the high accuracy of the method as well as its versatility, as it is readily applied to the above scenarios without any changes to the framework.
\end{abstract}



\begin{teaserfigure}
 \centering
    \includegraphics[width=\textwidth]{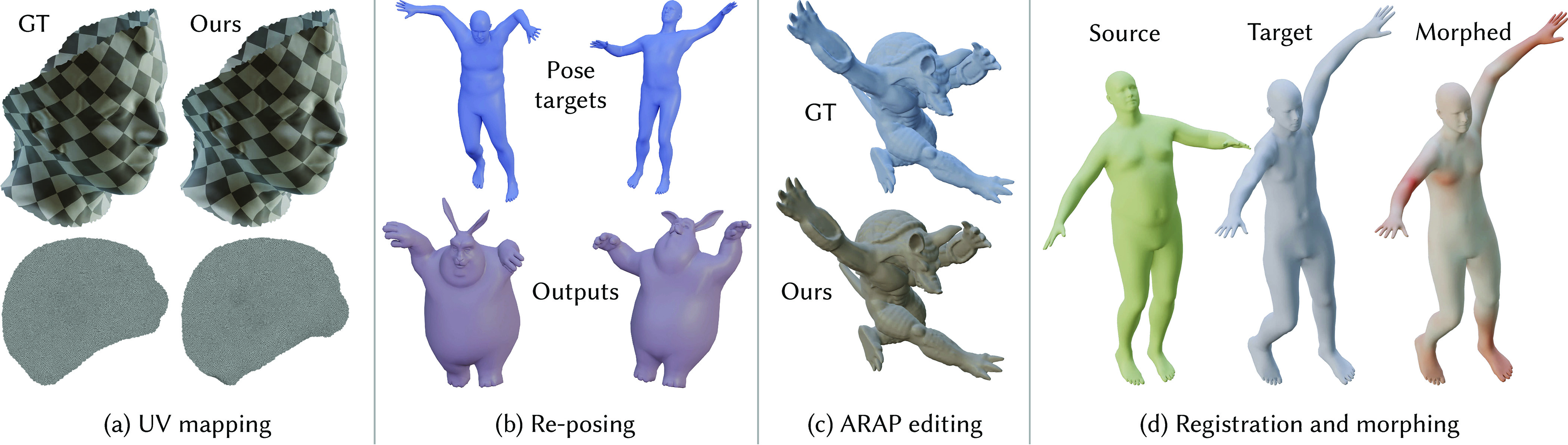}
    % \caption{\na{fix numbers}Pairs of parameterizations. The left one is produced by an optimization algorithm for parameterization, SLIM~\cite{Rabinovich:SLIM:2017}, and the right one is the prediction of our neural network, trained over examples parameterized by SLIM.  Results are shown on shapes unseen during training, with varying geometry, topology and discretizations. We report the Symmetric Dirichlet energy and the number of flipped triangle.  SLIM is an optimization algorithm and converges to a local minimum, hence our results exhibit higher distortion and more inverted elements than SLIM. However, our network runs at the price of a single feed forward and linear solve. }
    \caption{We present a versatile, detail-preserving and triangulation-agnostic neural framework for learning piecewise-linear mappings of meshes. Using the \emph{exact same core framework}, we show, from left to right: (a) learning to UV-map arbitrary surfaces; (b) re-posing a novel character (the bunny), unseen during training and with a completely different mesh topology, using a model trained only on humans; (c) learning to replicate As-Rigid-As-Possible~\cite{sorkine2007arap} deformations; and (d) morphing a human shape (green) to a target pose and body shape (blue), without input correspondences.}
    \label{fig:teaser}
\end{teaserfigure}

\maketitle

\section{Introduction}
%maps are important in GP
\label{sec:intro}
Computing mappings between 3D domains is a fundamental concept at the core of graphics and geometry processing, used in a wide range of generative and discriminative tasks, including 3D modeling, deformation, animation, UV-texturing, registration, correspondence-computation, remeshing, simulation, and fabrication.

Many such mapping tasks involve human priors which do not admit compact analytical representations, or otherwise require computationally-demanding numerical optimization at runtime. Hence, approaching such tasks in a data-driven manner, using predictive tools from modern statistical learning to accommodate semantic priors and approximate optimized solutions in a single forward pass, can have wide impact. Immediate potential applications include modifying 3D shapes to match a target such as an image, inferring high quality UV mappings from examples authored by artists, and integrating with direct optimization to speed-up tasks such as physical simulation.

In order to achieve accurate, high-quality results, it stands to reason to harness deep neural networks, which have proven immensely effective for complex regression tasks, and to learn mappings of 3D triangular meshes, which are arguably the most common 3D representation chosen by artists when fine detail and high accuracy are desired. This also has the benefit of directly integrating into the graphics pipeline, without the need to convert back and forth between different representations.

\begin{figure}
    \centering
    \includegraphics[width=\columnwidth]{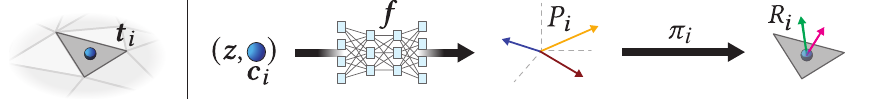}
    \caption{The neural aspect of our framework is summed up in applying an MLP to a single triangle at a time. The MLP, $\mlp$, receives as input the centroid feature $\centroid_i$ of the triangle $\tri_i$, along with the global code $\codeword$. It  outputs a matrix $\pred_i\in\Real^{3\times3}$, visualized as three vectors in $\Real^3$, which is then restricted to its action on the tangent space of the triangle $\tri_i$, through the restriction operator  $\proj_i$, to yield a matrix $\rest_i\in\Real^{3\times2}$, visualized as two vectors in $\Real^3$. }
    \label{fig:local_step}
    \vspace{-15pt}
\end{figure}

\vspace{-4pt}
\subsection{Challenges}
Inferring mappings of meshes with neural networks is still, to a large extent, an open problem. There are fundamental issues that make 3D surfaces (that is, 2-manifolds immersed in 3D) a harder domain to efficiently learn over than, e.g., 2D images. Methods attempting to predict mappings encounter two main challenges: \mbox{\textbf{(1)} 3D} surfaces, as opposed to the standard 2D pixel grid, have significant geometric and topological variation. There is no canonical way to associate corresponding points between a pair of surfaces, and the two instances may differ in their discretization (i.e., have different triangulations). Standard 2D convolutional architectures make assumptions about point ordering and parametrization that do not apply, in general, to collections of 3D surfaces. \mbox{\textbf{(2)} Mappings} of surfaces need to be \emph{detail preserving}. That is, fine geometric details should be preserved while mapping the coarse (lower-frequency) structure of the surface through a highly-nonlinear map.

Thus, previous learning-based mesh mapping methods devise specific architectures and mapping spaces to tackle one or both of these challenges. Some methods opt to work with one, fixed triangulation, which prevents their application to scenarios where the runtime triangulation is not given in advance, or where the training data themselves have diverse triangulations. Other methods define deformations over the ambient 3D space~\cite{jiang2020shapeflow}. However, ambient fields are not detail-preserving in general, and the burden falls on the network to ensure that the specific predicted fields protect mesh details.

Other methods choose a subset of ambient-space deformations which are detail-preserving by construction, e.g., deformations induced by sparse rigs such as cages~\cite{wang2019neural}. These methods are still not fully
shape-aware, but rather rely on the rig inducing a highly restricted deformation space. This in turn implies their success hinges on fitting the rig to the mesh to expose exactly the correct deformation subspace. Further, in cases where the rig is not applicable due to vast differences in shape or topology, the network cannot learn meaningful maps.

In order for a method to map meshes in a truly shape-aware manner, it needs to be defined intrinsically over the manifold surface itself. Non-ML mapping techniques commonly operate in the \emph{gradient domain} of the manifold, by considering the jacobians of the map w.r.t. the intrinsic gradient operator of the mesh. However, incorporating this technique into a neural network revives the first challenge of operating on heterogeneous, non-regular mesh topologies, as the intrinsic gradient operator is defined for each triangle w.r.t. its triangulation-specific tangent spaces.

\subsection{Approach}

In this work, we aim to tackle both challenges mentioned above, and devise a method for learning \emph{intrinsic, highly-accurate, detail-preserving} mappings, which is provably and experimentally shown to be \emph{triangulation-agnostic}.

Our method is based on the following straightforward observation: we can have a multi-layer perceptron (MLP) make a prediction of an \emph{extrinsic} field over ambient space, unaware of any triangulation, and then, for any desired mesh, we can use the mesh's differential operators to ``carve out'' an \emph{intrinsic} field of jacobians and compute a mapping of the given mesh from them.  By postponing any use of the mesh until \emph{after} the MLP prediction is made, we on one hand enable a mesh-aware, intrinsic loss that supports backpropagation, but on the other hand, prevent the MLP from any ability to use the mesh for inference, thereby forcing it to learn in a manner that cannot rely on a specific triangulation.

Concretely, our neural component (visualized in Figure \ref{fig:local_step}) comprises of the aforementioned MLP, which receives a single point in space as input and produces a $3\times 3$ matrix, thereby inducing a matrix field, in similar fashion to other geometric deep learning methods which make point-wise predictions~\cite{park2019deepsdf,groueix2018atlasnet}. We then define a simple restriction operator which restricts the \emph{extrinsic} field of linear transformations to an \emph{intrinsic} field of linear transformations, defined over the tangent bundle of the given mesh, representing the prediction of candidate jacobians. We further observe that the linear solve of a Poisson system, required to retrieve a continuous map from these projected pseudo-jacobians, can be implemented efficiently as a differentiable layer whose weights can be precomputed for each mesh. This enables us to train and perform inference of maps over a collection of meshes with varying triangulations, in spite of the Poisson linear system varying for each mesh in the dataset.

This construction achieves several goals simultaneously. (1) We make the framework fully agnostic to the triangulation, as the same trained network can be applied to a triangulation of any connectivity, and furthermore it is not biased by any triangulation-specific signal during training. (2) The MLP's prediction is interpreted intrinsically in the mesh's gradient domain, resulting in a detail-preserving, highly accurate map. Furthermore, since most mappings considered in practical applications vary gradually across the input shapes, their jacobians are low-frequency, smooth signals. As our MLP learns to map points in ambient space to the jacobians, it can exploit this smoothness and learn to reproduce them to high accuracy without exceeding its capacity. (3) Lastly, beyond mesh-agnosticism and accuracy, we also show our method is extremely versatile, as we do not change any aspect of the core framework (other than supplying it an additional input in the form of a pose vector for some of the human retargeting experiments) for any of the experiments in the paper.

We show these three properties of our framework -- accuracy, versatility, and triangulation-agnosticism -- by exhibiting its capabilities on a large variety of experiments, ranging from morphing, re-posing, reshaping, and deformation transfer of humans, to emulating physical simulations such as non-rigid deformations and collision handling. Lastly, we show that our method is, to the best of our knowledge, the first that can effectively learn to produce UV mappings by training over a heterogeneous collection of meshes.

\section{Related Work}
\label{sec:related}

\noindent \textbf{Map computation through jacobians.} Map computation is a keystone of many central areas of geometry processing and graphics, such as UV parameterization~\cite{Sheffer:meshparam:2007}, deformations induced from either physical simulation~\cite{Kim:DynamicDeformables:2020} or artistic manipulation~\cite{DeformationTutorial:2009}, registration~\cite{Sofien:2016}, and computing polycubes~\cite{Tarini:polycube:2004}. \change{Several geometry processing methods enable artist-driven mesh-editing through the gradient domain~\cite{yu2004mesh,lipman2004differential,sorkine2004laplacian}}. Furthermore, mappings are commonly defined by a variational principle as a minimizer of an energy, usually defined in terms of the jacobians, e.g., Dirichlet~\cite{Pinkall93computingdiscrete}, ARAP~\cite{Ligang:arap:2008}, LSCM~\cite{levy2002lscm}, and symmetric Dirichlet \cite{Smith:bijective:2015,Rabinovich:SLIM:2017}. Other methods aim to restrict the jacobians to ensure that they do not invert~\cite{schuller2013locally,du2020lifting}, or additionally bound the distortion of the jacobian~\cite{lipman2012bounded,myles2013controlled,aigerman2013injective,kovalsky2014controlling,weber2014locally}.
A central algorithm used in these methods is the block-descent local/global algorithm~\cite{liu2008local} which iteratively applies two steps, a local step that makes a per-triangle update to the jacobians into non-integrable matrices, followed by a global step which harnesses Poisson's equation to find the mapping which best-respects the local update. Our method can be viewed as an adaptation of a single iteration of this algorithm with a local, neural prediction step followed by solving Poisson's equation, along with a construction that enables us to apply it to a collection of meshes with different triangulations.

\noindent \textbf{Data-driven deformations.}
Deformations play a central role in animation, graphics and physics. Leveraging data-driven methods can enable automating deformation transfer~\cite{sumner2004deformation,gaovcgan2018}, inferring deformation spaces which  guide asset retrieval~\cite{uy2020deformation}, producing meshes composed of deformable parts~\cite{gao2019sdm}, use image input to reconstruct objects or deform templates~\cite{wang2018pixel2mesh,kanazawa2016learning}, and learn deformation subspaces without losing plausibility~\cite{holden2019subspace}.

\noindent \textbf{Rigged deformations.} In non-ML context, Skinning-based methods~\cite{skinningcourse:2014,Fulton:LSD:2018} use a rig with handles such as points~\cite{jacobson2011bounded}, skeletons~\cite{kavan2008geometric}, or enclosing cages~\cite{ju2005mean,lipman2008green}, which induce weights that propagate rig deformations to the underlying shape, thereby enabling its applicability to arbitrary meshes as well as point clouds. This process heavily relies on the accurate fitting of the rig, which is a subtle task. To automate it, heuristic-driven~\cite{Baran:rigging:2007} and neural techniques have been developed to predict and deform various rigs such as skeletal ones~\cite{AnimSkelVolNet,RigNet,Holden:inverse_rig:2015,li2021learning},  \change{point handles~\cite{liu2021deepmetahandles,jakab2021keypointdeformer}}, and cages~\cite{wang2019neural}. This approach can also be employed to use a human model as a rig which deforms their clothing~\cite{liu2019neuroskinning}.
 Inaccuracy in the rigging leads to severe artifacts, and the rig further usually restricts the architecture to a certain class of deformations, e.g., articulations for bones. None of these rigging methods is readily applicable to general mapping tasks, such as UV parameterization.

\noindent \textbf{Learning deformations of a single mesh.}
Many methods focus on a single mesh, or a set of variations of the same mesh~\cite{bogo2014faust,varol17_surreal,SMAL:2017,STAR:2020}, and define an arbitrary \emph{fixed correspondence} of vertices and/or faces to the entries of a predicted tensor, thereby enabling  machine learning algorithms to associate predicted quantities to geometric elements. This enables treating the input and output as general tensors of a homogeneous dataset and applying off-the-shelf tools for data analysis, such as Principal Component Analysis~\cite{anguelov2005scape}, Gaussian Mixture Models~\cite{Bogo:ECCV:2016}, as well as neural networks, which assign per-vertex coordinates~\cite{Shen:2021} or offsets from a simpler (e.g., linear) model~\cite{Bailey:2018:FDD,bailey2020fast,Romero:2021,Zheng:secondary_motion:2021,yin2021_3DStyleNet}. As they are not shape-aware in the sense discussed in Section \ref{sec:intro}, such methods often need to add additional regularizers such as ARAP~\cite{Sun:2021} or the Laplacian~\cite{cmrKanazawa18}. To get around the limitation to a single shape, some methods use a single template to deform partial parts of shapes \cite{litany2018deformable}. 

Similarly to us, \cite{tan2018meshvae,gaovcgan2018} also consider jacobians of deformations. They   propose to autoencode features extracted from jacobians in order to learn a latent deformation space. However, unlike us, they consider a single mesh, with a fixed assignment of the predicted tensor's entries to the mesh's elements. Thus, they are restricted to only train on this one mesh's triangulation, and cannot be applied to a collection unless the triangulations are in one-to-one correspondence. In contrast, our framework predicts continuous jacobian fields, and thus is readily applicable to heterogeneous collections with diverse triangulations. Furthermore, in Section \ref{sec:comparisons} we show that even when restricted to a single mesh, predicting jacobian fields significantly outperforms the fixed-assignment approach.

\noindent \textbf{Discretization agnostic representations.} Many works seek to avoid the discrete nature of meshes. Deformation fields are commonly used to generalize across different triangulations and even geometric domains. Specifically, one can learn to predict an implicit vector field which maps every point in the volume to its new location~\cite{groueix2018coded,Groueix19,Yang:2021,jiang2020shapeflow,huang2020meshode}. This representation acts in a point-wise manner, and is not aware of the underlying surface,  thus these maps tend to not preserve local surface details. By operating in the gradient domain our method is detail-preserving by construction. 
Other methods avoid meshes by defining the surface itself as a mapping of a plane through a neural network~\cite{Groueix19,williams2019deep,morreale2021neural}, or an SDF~\cite{park2019deepsdf},  however their implicit and non-discrete representation makes it non trivial to manipulate them accurately. In the context of deep-learning, Atlasnet-like works explored leveraging jacobians~\cite{bednarik2020shape}, however this is to  regulate their atlases and does not enable defining mappings of these surfaces.  Networks that are agnostic to the discretization of surfaces have recently been studied in \cite{sharp2022diffusionnet}, however that work is focused on analysis of the surface by designing discretization-invariant diffusion operators, and less fitting for producing deformations of high-resolution surfaces (see also Fig \ref{fig:diffnet} for a comparison of the two methods).

\section{Approach}

We begin describing our method by first elaborating on the main algorithm of our framework, which enables us to predict shape-aware piecewise-linear maps, in a manner completely agnostic to triangulation.

At the prediction stage, we are given as input a  mesh  $\source$, and  a \emph{global code} $\codeword$, which is a vector that is compiled differently for each of the experiments -- we elaborate on its computation in Section \ref{s:experiments}. As an example, $\codeword$ can be the concatenation of the angles at the joints of a skeleton, encoding the desired position to which a given mesh of human should bend. We denote the  ground truth map (e.g., the correct mapping of the rest pose human to the desired pose) as $\gtmap$. 

Our goal is to design a network that, conditioned on  $\codeword$, will predict a map $\predmap$ that matches the ground truth $\gtmap$. Furthermore, we set it as our goal to design this network in a way that will enable it to learn and perform inference in  complete agnosticism to the source domain's triangulation, but that will still produce a shape-aware, detail-preserving map  through the gradient domain of the mesh.

 We progress in a bottom-up manner, beginning with a brief review of a few basic concepts and definitions of piecewise-linear mappings needed for our construction, then move on to describing our neural architecture for predicting mesh-agnostic piecewise linear maps, and finish with a discussion on how to train it on a  dataset of mappings. 
\subsection{Preliminaries}
\label{ss:pre}
We assume the mesh $\source$ is a  2-manifold triangular mesh of a single connected component, embedded in $\Real^3$, with vertices $\vertices$ and triangles $\triangles$.
The \emph{tangent space} at a triangle $\tri_i\in\triangles$
is the linear space orthogonal to its normal, denoted $\tangent_i$. We choose two column vectors which form an oriented orthonormal basis to the tangent space, which we call a \emph{frame}, $\fram_i\in\Real^{3\times2}$ of the triangle.
\paragraph{Piecewise linear maps}  
As is common in geometry processing, we focus on the space of \emph{piecewise-linear mappings} of a given mesh, meaning that the restriction of the map $\predmap$ to any triangle $\tri_i$, denoted $\predmap|_{\tri_i}$, is affine, i.e., is the sum of a linear map and a constant translation. This is arguably the most common family of maps used when considering mappings of meshes. A piecewise linear mapping $\predmap$ of a mesh can be uniquely defined by assigning a new position to each one of the vertices, $\vertices_i \to \predmap_i$. The mapping is then well-defined also for any point inside a triangle by linearly interpolating the map from the vertices. From now on, we abuse notation and interchangeably  use $\predmap_i$ to denote the $i$'th vertex's new position, and consider $\predmap$ as a matrix of same dimensions as $\vertices$.
\paragraph{Jacobians.} Analogously to the smooth setting, the jacobian at triangle $\tri_i$  of the map $\predmap$ is a linear transformation of dimensions $3\times2$ from the triangle's tangent space to $\Real^3$, denoted $\jac_i\parr{x}:\tangent_i \to \Real^3$, defined as the linear part of the map restricted to that triangle,  $\predmap|_{\t_i}$.

For a given map $\predmap$, the jacobian $\jac_i$ can be explicitly computed as a $3\times2$ matrix in the coordinates of the frame $\fram_i$ by solving 
\begin{equation}
\jac_i \fram_i^T\brac{v_k - v_j,v_l - v_j} =  \brac{\phi_k- \phi_j,\phi_l - \phi_j}
\end{equation}
where $v_j,v_k,v_l$ are the triangle's vertices, and $\phi_j,\phi_k,\phi_l$ are their images under $\predmap$.  
Solving the above linear system yields a linear operator denoted $\nabla_i$, that maps $\predmap$ to the jacobian $\jac_i$ in the basis $\fram_i$:
\begin{equation}
\label{eq:jac}
\jac_i  =  \predmap\nabla_i^T.
\end{equation}
 $\nabla_i$ is defined as the gradient operator of triangle $\tri_i$, expressed in the basis $\fram_i$. The gradient operator $\nabla$ of the mesh is thus defined as the linear operator mapping $\Phi$ to the stack of all jacobians.

\paragraph{Poisson equation.}  Given an arbitrary assignment of a matrix $M_i\in\Real^{3\times2}$ to each triangle, one can retrieve the map $\predmap^*$ whose jacobians $\jac_i=\predmap^*\nabla_i^T$ are closest to $M_i$ in the least-squares sense, by solving the widely-used \emph{Poisson equation},
\begin{equation}
\label{eq:lsq_poisson}
\predmap^*  = \min_{\predmap} \sum{\abs{\tri_i}\norm{\predmap\nabla_i^T - M_i}^2},
\end{equation}
where $\abs{\tri_i}$ is the area of the triangle $\tri_i$ on the source mesh $\source$.

The solution is obtained by solving the linear system
\begin{equation}
\label{eq:poisson}
 \predmap^* = L^{-1}\mathcal{A}{\nabla^T} M,   
\end{equation}
 where $\mathcal{A}$ is the mesh's mass matrix, $L = \nabla^T \mathcal{A} \nabla$ is the mesh's cotangent Laplacian, and $M$ is the stacking of the input matrices $M_i$. The solution is well-defined up to a global translation which can be resolved by setting, e.g., $\predmap^*_0 = \vec{0}$ and modifying the above equations accordingly.

\subsection{Predicting Triangulation-Agnostic, Intrinsic Mappings}

\begin{figure*}[t]
    \centering
    \includegraphics[width=\textwidth]{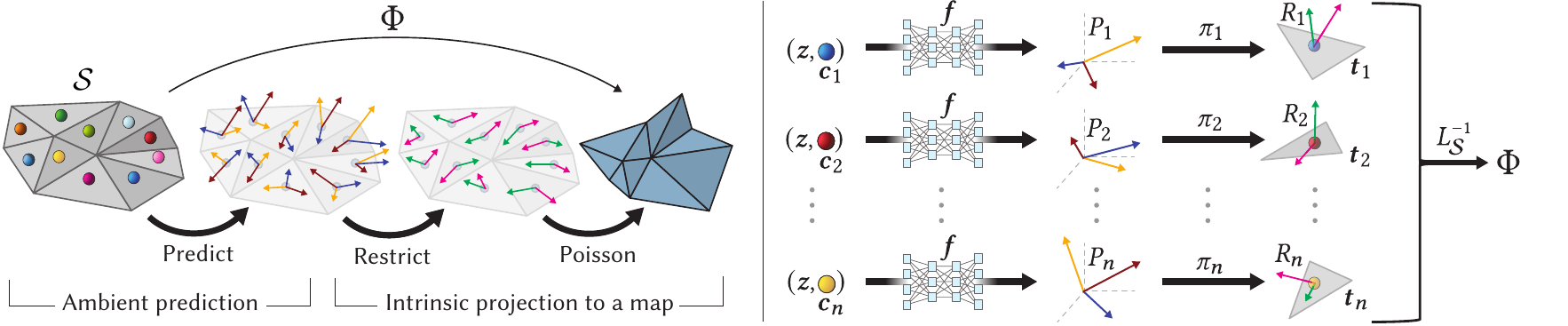}
    \caption{The main inference algorithm for predicting mappings,  depicted once through its geometric action (left) and once through its dataflow (right).  For a given mesh $\source$ and a global code $\codeword$, the centroid feature $\centroid_i$ of each triangle $\tri_i$ of the mesh is concatenated to $\codeword$ and plugged into the MLP $\mlp$ in parallel. The MLP predicts a matrix  $\pred_i\in \Real^{3\times3}$ thereby inducing a field of matrices over all triangles. The matrices are then \emph{restricted}, in parallel, via $\proj_i$ to the tangent spaces of their respective triangles, yielding $\rest_i\in\Real^{3\times2}$. $\rest_i$ are then plugged to the Poisson solve which computes a map $\Phi$, by solving a linear equation w.r.t. the given mesh's own Laplacian $L_\source^{-1}$,  intrinsically projecting the network's prediction to a mesh-specific, detail-preserving mapping $\predmap$ of the mesh.}
    \label{fig:pipeline}
\end{figure*}
We now begin elaborating on our framework. See Figure \ref{fig:pipeline} for the detailed pipeline.
 Intuitively, one may find  analogies between the prediction algorithm and one, \emph{single} iteration of the local-global algorithm widely used in geometry processing, e.g., ARAP~\cite{liu2008local}, with our custom local step (shown in Figure \ref{fig:local_step})  including an MLP and a restriction operator from ambient space into the triangle's tangent space.

The design of our framework should posses two supposedly-conflicting properties: 1) it should be completely agnostic to the triangulation of the meshes, and neither the framework's input, output, nor structure should  be modeled with respect to a specific triangulation; 2) it should predict \emph{intrinsic}, gradient-domain mappings so as to accurately preserve details. However, working in the gradient domain entails working in the tangent spaces $\tangent_i$ using gradient operators $\nabla_i$, which are \emph{triangle specific}, entailing a naive approach will couple the method to a specific triangulation.

We address both these challenges in an algorithm consisting of two stages:  1) an MLP, conditioned on the global code $\codeword$, predicts a   $3\times3$ matrix for any given point $p$, thereby inducing an ambient field of extrinsic matrices. 2) The extrinsic matrix field is reinterpreted and restricted by mesh-specific, basic linear-algebra operations to an \emph{intrinsic} field of $3\times2$ matrices defined over the tangent spaces $\tangent_i$ of the mesh. The intrinsic matrices are input to the Poisson solve \eqref{eq:poisson} of the mesh, implemented as a differentiable layer over the GPU, yielding  a mapping of the mesh through the gradient domain. 

Intuitively, the MLP makes its prediction at the end of step 1, and receives only a single triangle centroid each time, thus it cannot make any global inference regarding the triangulation itself based on the input. The only global inference stems from the concatenated code $\codeword$, which in all our experiments does not contain any information about the triangulation itself, rather solely the shape the triangulation represents, ensuring the MLP is blind to the triangulation, thereby making the framework  \emph{triangulation agnostic}. Step 2) which transpires outside of the neural component,  uses the specific mesh's differential operators along with simple linear algebra to pull the prediction of step 1) into the specific gradient domain of the specific mesh, thereby yielding a shape-aware intrinsic mapping, as desired.
Next, we elaborate on each step in sequence.

\textbf{Step 1: Predicting matrices in $\Real^{3\times3}$.}
In similar fashion to other networks that make per-point  predictions~\cite{park2019deepsdf,groueix2018atlasnet} we use an MLP $\mlp$ which receives as input a single point $p$, concatenated to the global code $\codeword$, and outputs a $3\times3$ real matrix, thereby inducing a field of matrices over ambient space and not tied to a specific mesh.  Given a specific mesh $\source$ to deform, independently for each triangle $\tri_i$, we feed each of its centroids $\centroid_i$ as the point $p$, along with the global code $\codeword$, thereby assigning a  matrix $\pred_i\in\Real^{3\times3}$ to the triangle $\tri_i$,
\begin{equation}
\label{eq:pred}
     \pred_i = \mlp\parr{\codeword,\centroid_i} \in\Real^{3\times3}.
\end{equation}
 Each centroid $\centroid_i$ is represented as a fixed, precomputed vector, consisting of a concatenation of the centroid's 3D position, its normal, and a Wave-Kernel signature~\cite{aubry2011wave}. For performance gain, we can of course apply the MLP in parallel on a large batch of centroids over the GPU. As our goal is to define an intrinsic jacobian for each triangle, we next \emph{restrict} each extrinsic matrix to its action on the tangent space of its corresponding triangle.

\textbf{Step 2: Compute a map from the extrinsic field, using the mesh's intrinsic differential operators.}   The predicted extrinsic matrix $\pred_i$ of triangle $\tri_i$ can be projected to an intrinsic linear map, by considering its \emph{restriction} to the subspace of  $\tangent_i$, expressed in the frame $\fram_i$:
\begin{equation}
\label{eq:rest}
    \proj_i\parr{\pred_i} \triangleq \pred_i \fram_i.
\end{equation} 
We denote $\pred_i$'s  restriction as $\rest_i\in \Real^{3 \times 2}$.

With the restricted matrices $\set{\rest_i}$ at hand, to produce the final map, we plug them into the Poisson system, Eq. \eqref{eq:poisson}, defined via the Laplacian and gradient operator of $\source$. Fortunately, we observe that the Poisson solve can be implemented as a custom differentiable layer, as Equation \eqref{eq:poisson} represents a linear transformation, and hence when back-propagating the incoming gradient $g$, the back-propagation from this linear layer amounts to solving \eqref{eq:poisson}
 again for a different right hand side input, defined by the gradient. 
 
 To solve Equation \eqref{eq:poisson} rapidly, as is desired during consecutive evaluations (e.g., during training or inference), we follow the common practice of computing a decomposition of $L$ into two tridiagonal systems in advance (we use an LU decomposition), during  the preprocessing of the data before training.  Then, during training or evaluation, the Poisson solve can be quickly obtained by loading the decomposition into GPU memory, and performing  backsubstitution on the two tridiagonal systems, on the GPU.

We summarize the inference algorithm in Algorithm \ref{alg:infer}  and the preprocessing procedure in Algorithm \ref{alg:pre}.
 \begin{algorithm}
\caption{Inference}\label{alg:infer}
\begin{algorithmic}
\State \textbf{Input:}  codeword $\codeword$, source mesh $\source$,  $\nabla_\source$, LU decomposition of  $L_\source$ 
\For{ each centroid $\centroid_i\in\mathcal{C}$, in parallel on the GPU,} 
\State \indent Apply $\mlp$ to concatenation of $\parr{\codeword,\centroid_i}$ to get $\pred_i$ via Eq. \eqref{eq:pred}.
\State \indent Restrict $\pred_i$ to the tangent space   via Eq. \eqref{eq:rest} to get $\rest_i$.
\EndFor
\State Compute $\predmap$ via Eq. \eqref{eq:poisson} using $\nabla_\source$ and the LU decomposition of $L_\source$.
\State \textbf{Output:} the map $\predmap$ assigning a new position to each vertex of $\source$.
\end{algorithmic}
 \end{algorithm}
 \begin{algorithm}
\caption{Mesh Preprocessing}\label{alg:pre}
\begin{algorithmic}
\State \textbf{Input:} Mesh $\source$ to preprocess.
\State Compute and store the centroid features  $\mathcal{C}_\source = \set{\centroid_i}$.
\State Compute and store the frames $\fram_\source = \set{\fram_i}$.
\State Compute and store the gradient operator $\nabla_\source = \set{\nabla_i}$.
\State Compute and store the Laplacian $L_\source$ and its LU-decomposition.
\end{algorithmic}

 \end{algorithm}
 
 For a collection of meshes, we run Algorithm \ref{alg:pre} on each one of them in advance and store all computed data. Then, during training or inference, for each mesh, we load its data and run Algorithm \ref{alg:infer}.

 \subsection{Training}

\begin{figure}[t]
    \centering
    \includegraphics[width=\columnwidth]{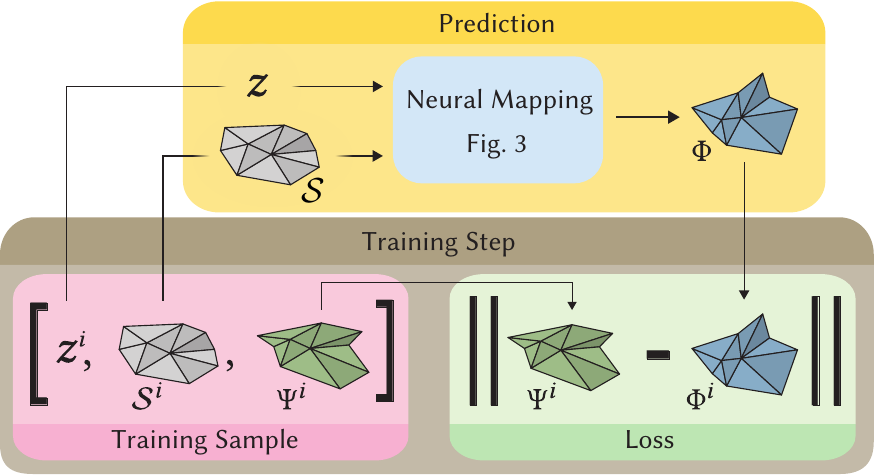}
    \caption{Top: at inference, the network outputs a mapping $\predmap$ of a mesh $\source$ given as input, conditioned on the global code $\codeword$. Bottom: during training, we iterate over triplets, each consisting of a mesh $\source^i$, groundtruth mapping $\gtmap^i$, and the associated global code $\codeword^i$, and predict a mapping $\predmap^i$, which is then compared with respect to the ground truth through the $L2$ distance between their vertices and their jacobians. Each training or evaluation sample may have a completely different triangulation without affecting the prediction.}
   \label{fig:training}
\end{figure}

 Figure \ref{fig:training}, bottom, shows the training procedure of the framework. The training is conducted over a dataset of maps, defined via triplets, $\set{\parr{\source^i,\gtmap^i,\codeword^i}}_{i=1}^n$ comprised of a mesh, its corresponding ground-truth mapping, and the global code on which the prediction of the network should be conditioned. In most experiments, $\codeword^i$ is obtained by a PointNet encoder which is trained in tandem with the deformation network, however any encoder or shape representation can be used.  During training, we iterate over the triplets and for each triplet, we train the network to predict a map $\predmap^i$ of the mesh $\source^i$, conditioned on the global code $\codeword^i$, using a loss defined w.r.t. the ground-truth map, $\gtmap^i$. 
 
 \paragraph{Losses. }We optimize for two losses: first, the vertex-vertex loss between the prediction $\predmap$ and the ground truth map $\gtmap$,  
 \begin{equation}
     \vloss = \sum{\abs{\vertices_j}\norm{\predmap_j - \gtmap_j}^2},
 \end{equation}
 where $\abs{\vertices_i}$ is the lumped mass around the $i$'th vertex in $\source$. Since the network's prediction is well-defined up to a global translation (see last paragraph of Section \ref{ss:pre}), we shift both the prediction and the ground truth so that their meshes' center of mass is at the origin, before computing this loss.
 
 We also measure the difference between the restricted predictions $\set{\rest_i}$  and the ground truth jacobians $\jac_i=\set{\gtmap\nabla_i^T}$,
 
 \begin{equation}
 \label{eq:jloss}
     \jloss = \sum{\abs{t_j}\norm{\rest_j-\jac_j}^2}.
 \end{equation}

 Our total loss is
 \begin{equation}
 \label{eq:tot_loss}
     \totloss = 10\cdot\vloss+\jloss.
 \end{equation}
 We use this loss in all experiments. Note that both losses are achieved through the use of the specific mesh's differential operators,  after inference. 
We summarize the training in Algorithm \ref{alg:train}.  
 \begin{algorithm}
\caption{Training epoch}\label{alg:train}
\begin{algorithmic}
\For{each mesh and ground-truth map $\source,\gtmap$ in the dataset}
\State Load the LU decomposition and $\nabla$ of $\source$ to GPU memory.
        \State Encode $\source,\gtmap$ into a codeword $\codeword$. \Comment{experiment-specific}
        \State Input $\source$ and $\codeword$ to Algorithm \ref{alg:infer} to compute the mapping $\predmap$. 
        \State Compute the loss $\totloss$, Eq. \eqref{eq:tot_loss}.
        \State Optimize parameters of $\mlp$ and  encoder via back-propagation.
      \EndFor
\end{algorithmic}
 \end{algorithm}

 \paragraph{Encoding the global code $\codeword$ and the centroids.} The global code $\codeword$ is constructed differently in each of our experiments and we detail it for each of them in Section \ref{s:experiments}. At a high level, we use two types of encodings: First, we use fixed, pre-given dataset-specific parameters (such as SMPL pose parameters); second, we use Pointnet~\cite{qi2017pointnet} when there's need to encode the shape of either $\source$, or $\gtmap$, or both. In those cases, we sample $1024$ points on the mesh and compute a Wave-Kernel Signature~\cite{aubry2011wave} of size 50 for each point, and feed those along with the points' 3D position and normal into pointnet, which we train along with the MLP $\mlp$. Similarly, for  each centroid of a triangle fed into the MLP, we attach its 3D coordinates, normal, and a Wave-Kernel Signature of size $50$ as one vector. 

\subsection{Discussion: restriction through the frames $\set{\fram_i}$}

\begin{figure*}
    \centering
    \includegraphics[width=\textwidth]{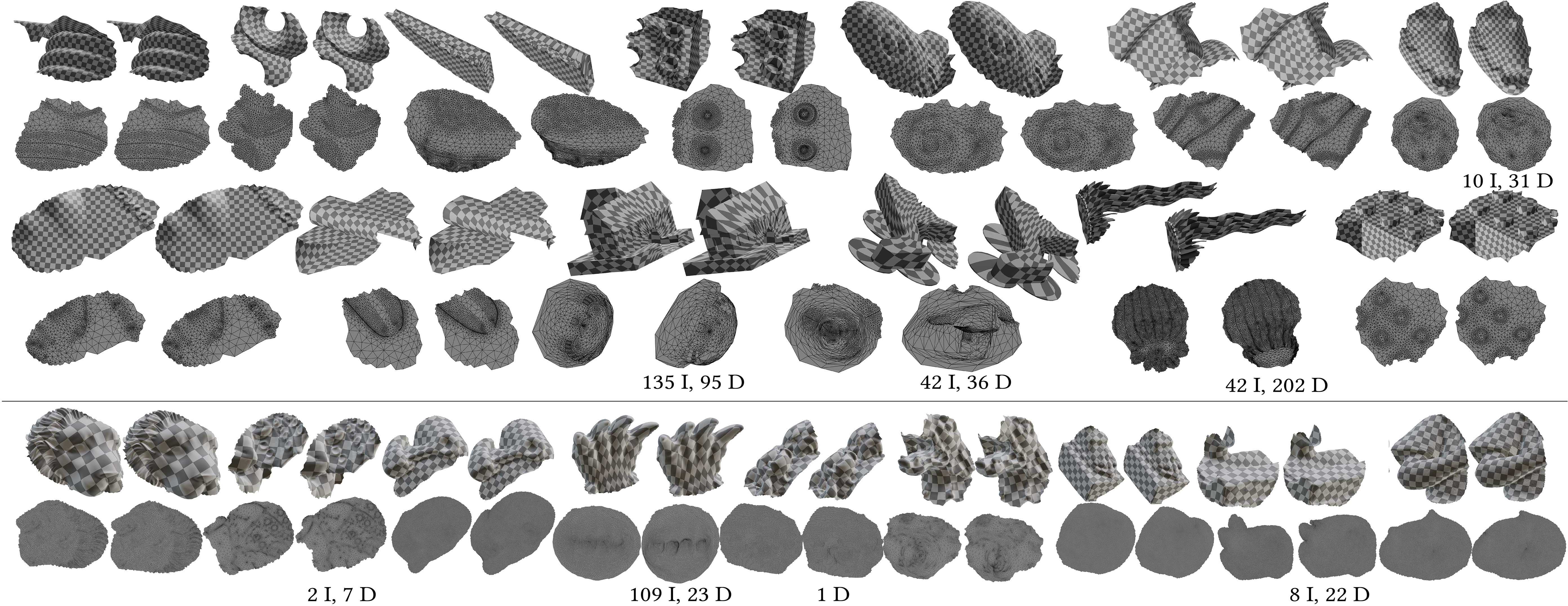}
    \caption{UV maps generated by our network (right in each pair), compared to SLIM~\cite{Rabinovich:SLIM:2017} (left in each pair). For each example we show the 3D model textured using the UV map, and the 2D UV map below it. In case there were inverted elements, we report their number next to ``I", and in case there were triangles with high ($>10$) distortion, we report their number next to ``D". Top: results on the test set from the dataset generated for this experiment,  of patches extracted from Thingi10K~\cite{zhou2016thingi10k}.  Bottom row: evaluation on a set of meshes generated from other datasets than Thingi10K.}
    \label{fig:uv_all_in_one}
\end{figure*}

For a tangent space $\tangent_i$, there exist infinitely-many orthonormal bases, and we choose one arbitrarily. The frame $\fram_i$ changes the representation both of the ground-truth jacobians as well as the predicted $\rest_i$, and hence, our framework has to be  invariant to the choice of $\fram_i$, in order to ensure it does not learn a frame-specific representation which would strictly prevent it from generalizing and learning over collections of meshes. Therefore,  we provide a trivial proof in Appendix \ref{a:lemma} showing that our framework is completely invariant to the choice of frames.\begin{claim}
Our framework is completely invariant to the choice of frames $\set{\fram_i}.$
\end{claim}

Additionally, we note that the representation of $\rest_i$ and the jacobians in frames in the tangent space follows the standard intrinsic definition used in differential geometry. However, the intrinsic restriction operation $\proj_i$ could alternatively be described extrinsically, on $3\times3$ matrices, by keeping $\rest_i$ in world coordinates and simply nullifying its action on the triangle's normal, $\rest_i = \pred_i\parr{I - \vec{n_i}\vec{n_i}^T}$. Both restriction approaches are comparable in the time it takes to perform them. However, in this extrinsic approach the restricted matrix would be a matrix with $9$ entries, and  a singular value of $0$, while in the intrinsic approach the restriction has $6$ entries and full rank. We choose the more compact and stable representation.

\subsection{Technical Details}

We use a $5$-layer fully-connected MLP with ReLU activation and group norm~\cite{wu2018group} after each layer. Hidden layers are of size $128$, with the first layer's input dimensions depending on the size of $\codeword$, and the last layer's output being the $9$ entries of $\pred_i$. We add the identity matrix to the prediction so that when the MLP outputs the zero matrix the prediction is the identity. 

 The only encoding we use except for raw parameters like a human's pose parameters, is a PointNet~\cite{qi2017pointnet} encoding of the given shapes, which receives $1024$ points sampled uniformally on the mesh, along with their normals and Wave-Kernel Signatures~\cite{aubry2011wave} of size $50$. We modify PointNet to use group norms as well. Both the MLP and Pointnet are trained simultaneously.

We use PyTorch~\cite{NEURIPS2019_9015}, along with CuPy~\cite{cupy_learningsys2017} and \emph{torch-sparse} which are needed to represent the various differential operators and sparse matrices on the GPU. When the dataset we train on encompasses a one-to-many mapping (e.g, when mapping an SMPL model to multiple poses), we train on batches (usually of size $32$) that include a single source mesh and multiple target mappings, thereby reusing the differential operators and performing the LU-solve over a batch. We train using the Adam~\cite{KingmaAdam} optimizer, which we initialize with a learning rate of $10^{-3}$. Once plateaued, we reduce the learning rate to $10^{-4}$, and train until plateauing again. 

While we cache the LU-decomposition during preprocessing, we note that the computational overhead of the LU decomposition is within reason for performing it on-the-fly during training and evaluation. This could be useful in scenarios in which the source mesh is assumed to dynamically change during training, such as within a remeshing framework.

Lastly, we note that while symmetric matrices are usually factorized via an LDL decomposition, we could not find an implementation that fit into our differentiable framework, and thus opted to use an LU decomposition using SciPy's SuperLU decomposition.

\section{Results}

We now turn to evaluate our framework's efficacy in various applications and tests, designed to exhibit its three key traits, namely: detail preservation, mesh-agnosticism, and versatility as a general framework for learning mappings. To show the last trait, we opt to evaluate our framework on a broad range of scenarios, and do not introduce  \emph{any} custom modification for any of these experiments, beyond changing the input encoding. We further emphasize that to encode the shapes we solely use a PointNet encoder which we do not consider part of our contribution, as we focus on our ability to decode mappings and not on encoding.

\paragraph{Metrics} We measure error via the L2 distance  between a vertex and its ground truth position, after normalizing both the predicted mesh and ground-truth mesh to the unit sphere. We report in  Table~\ref{tab:quantitative_results} the average L2 distance summed over vertices and shapes in the dataset. We also report the L2 distance between predicted and ground-truth jacobians, as well as the average angular error on the normals.

\label{s:experiments}
\subsection{Learning UV parameterizations}

Computing UV parameterizations is a fundamental task within the graphics pipeline which, to the best of our knowledge, has not been explored successfully by deep-learning techniques as of now. This in large part is due to this task requiring very high accuracy, as a wrong prediction can cause the embedding to self-overlap, and even a slight perturbation can highly-distort triangles or invert their orientation. Furthermore, this is a task only feasible by a framework which can operate on arbitrary triangulations. Both these hurdles are fit for our method to overcome.

Since a large dataset of artist-authored UV's is not publicly available, we opt to evaluate our network's ability to emulate UV maps computed by an optimization-based algorithm, Scalable Locally Injective Maps~\cite{Rabinovich:SLIM:2017} (SLIM). SLIM is a state-of-the-art algorithm, targeting low distortion, and is guaranteed to output a UV parameterization that has no triangles with inverted orientation, making it an extremely challenging target to replicate via learning.

We created a parameterization benchmark by randomly extracting $100K$ disk-topology patches (SLIM requires disk-topology) by iterating over the  meshes in the Thingi10K~\cite{zhou2016thingi10k} dataset, sampling random points and growing regions from them to random radii. We then parameterize the patches using SLIM. Since the UV parameterization is invariant to rigid motions of the 3D patch, during data preparation we align the patch to the 2D plane s.t. the   distance of its vertices to the plane is minimized. The UV map itself is well-defined up to a rigid motion in the plane,  and hence we align it to the patch by solving a 2D procrustes problem which rigidly shifts the UV map. We then train our network by feeding it a global shape description of the patch via a PointNet encoding, and train with the loss $\totloss$. In this experiment, we treat the output as $2$-dimensional and ignore the output $z$-coordinate.

\begin{table}[t]
\centering
\small
% \resizebox{\linewidth}{!}{
\begin{tabular}{lcc|cc}
\toprule
     &  \multicolumn{2}{c}{Distorsion} & \multicolumn{2}{c}{Flips}   \\
   $\downarrow$  &   avg. D>10 & med. D>10 &  avg. \#I   & \#I>0\ \%    \\
\midrule
    \textbf{Thingi10K} & &  \\
    \textit{\ \ \ SLIM (GT)}  &   0.04 & 0.0 &  0.0 & 0   \\
    \textit{\ \ \ Ours}  &  5.67 & 0.0 & 3.44 & 19  \\
    \textit{\ \ \ \cite{sharp2022diffusionnet}} & 158.7 & 132.0 & 296.3 &  99.9    \\
    \textbf{UV-generalization} & &  \\
    \textit{\ \ \ SLIM (GT)} &  0.00 & 0.0& 0.0 & 0   \\
    \textit{\ \ \ Ours} &   1.99 & 0.0 & 0.774 & 10    \\ 
    \textit{\ \ \ \cite{sharp2022diffusionnet}} & 590.0 & 513.0 & 866.5 &  100.0\\
\bottomrule
\end{tabular}
% }
\caption{
{
\textbf{UV Distortion.}  We report the average (\textbf{avg. D>10}) and median (\textbf{med. D>10}) number of triangles per mesh whose distortion is higher than $10$, as well as the average number of flipped triangles per mesh (\textbf{\#I}), and the percentage of meshes with at-least one flipped triangle (\textbf{\#I>0} \%). We report these statistics for our prediction, the ground truth UVs generated by SLIM~\cite{Rabinovich:SLIM:2017}, \change{and for DiffusionNet~\cite{sharp2022diffusionnet}}. We show results over the test split from the Thingi10K~\cite{zhou2016thingi10k} dataset which was generated for training, as well as on a dataset composed of meshes that are not part of the Thingi10K dataset. }
}
\label{tab:uv}
\end{table}

Table \ref{tab:uv} details statistics on the UV parameterization' quality. We measure the parameterization's distortion via $\text{max}(\sigma_1,1/\sigma_2)$, where $\sigma_1\geq\sigma_2$ are the singular values of the jacobian. Figure \ref{fig:uv_all_in_one} displays a qualitative comparison of the network's prediction with the ground truth results of SLIM. The top two rows show results on the test set from Thingi10K, while the bottom row shows results on patches extracted from meshes retrieved from other repositories. We denote by \textbf{I} and \textbf{D} the number of inverted elements and ones with distortion higher than $10$, when they exist. 

As SLIM is an optimization algorithm attaining a minimum of the symmetric Dirichlet energy, it is impossible for us to exceed its performance in the metrics it is optimized for. Nonetheless, we note that our predictions closely-resemble those of SLIM's, and in many cases we manage to yield viable parameterizations with no inversions and possessing relatively-low distortion. 

In the bottom row of Figure \ref{fig:uv_all_in_one}, we further evaluate our network's generalization capabilities by mapping patches extracted from models outside of the Thingi10K dataset, with distinctly different features. The quantitative results in Table \ref{tab:uv} prove to be better than those over the Thingi10k test set, but this is mainly due to the two datasets varying in their nature (e.g., more mechanical parts in Thingi10k). Attempting to go beyond this level  of detail, or to meshes with highly extruding parts such as hands, proves to be beyond the generalization capabilities of our network which produces highly self-overlapping parameterizations.

Of course, for high-quality production assets, artifacts such as inverted elements are strictly not acceptable. However, we believe that this is a first important step towards applying learning algorithms to the UV-mapping problem. 
We also note that the trained network can be incorporated as a differentiable layer in a deep-learning pipeline, to incorporate UV-dependent losses in order to, e.g., guide segmentation.

\subsection{Re-posing and Registration}
\label{subsec:re-posing_and_registration}
\begin{figure}[t]
    \centering
    \includegraphics[width=\columnwidth]{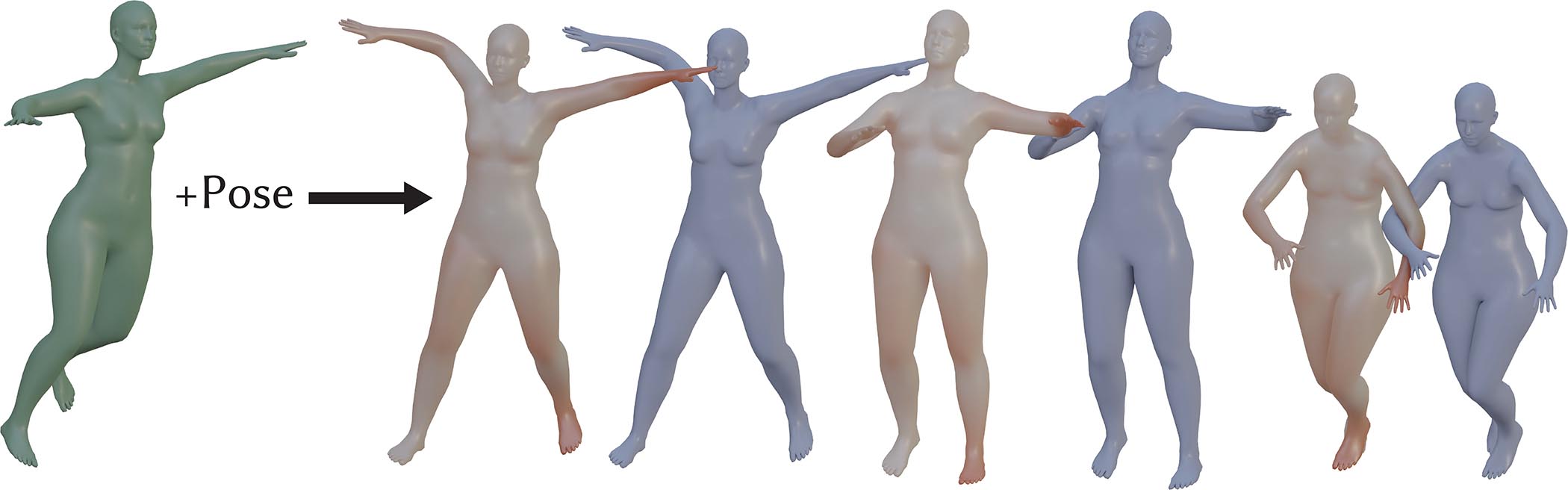}
    \vspace{-15pt}\caption{\textbf{Learning to re-pose humans.} We learn the STAR~\cite{STAR:2020} dataset's pose parameter space: the network is given a mesh of an arbitrary human (left, green) in an arbitrary pose, along with the parameters of the pose it should be moved to, and re-poses it correctly. Ground truth colored in blue, and predictions w.r.t. error (reddest being $0.05$).}
    \label{fig:pose_params}
\end{figure}
\begin{figure*}
    \centering
    \includegraphics[width=\textwidth]{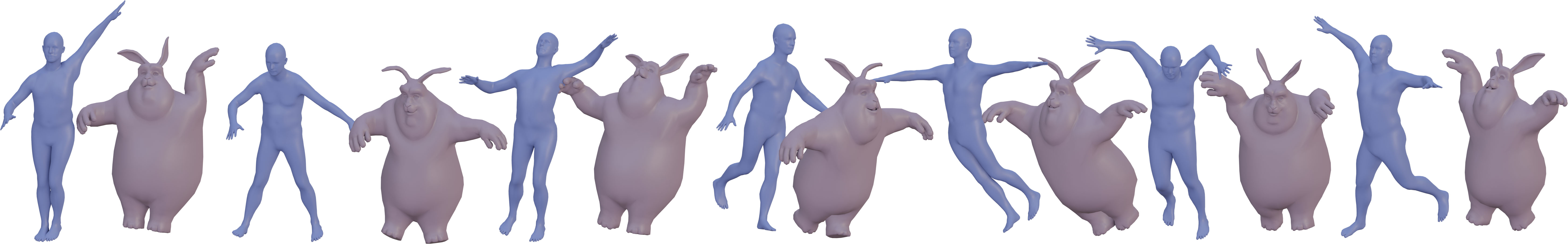}
    \caption{\textbf{Generalization of the re-posing network to Big Buck Bunny.} Our network was trained solely on training samples of human meshes from the STAR~\cite{STAR:2020} dataset, and is applied here to the bunny -- an unseen mesh, with different triangulation and geometry. We show a STAR mesh in blue, in the ground-truth pose, and the network's reposing of the bunny. The training set consists of one fixed triangulation, which is different than that of the bunny, hence this result can only be achieved by a triangulation-agnostic framework.}
    \label{fig:bunny}
\end{figure*}
\begin{figure*}
    \centering
    \includegraphics[width=\textwidth]{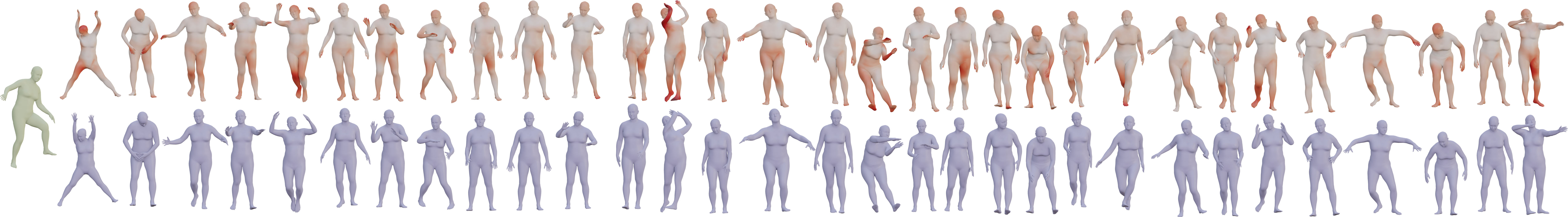}
    \caption{\textbf{Learning to register humans.} Our network  learns to correctly register any input random SMPL~\cite{Bogo:ECCV:2016}  model to itself in arbitrary poses. Source in green, targets in blue, and the deformation colored based on error. }
    \label{fig:smpl}
\end{figure*}

Computing deformations of human characters is a challenging task, which demands the computed maps to be highly accurate to capture articulated deformations correctly, while at the same time preserve the fine features of the humans. As in all other experiments, we  use a PointNet encoding for capturing the shape of the various humans, and do not modify the network to accommodate specifically for human anatomy in any way.

For this experiment, we leverage two datasets, SMPL~\cite{Bogo:ECCV:2016, varol17_surreal} and STAR~\cite{STAR:2020}. Both datasets comprise of a parametric model, which receives shape parameters controlling the appearance (e.g., tall or short) of the human, as well as pose parameters, which define the pose the articulated human assumes. We use these parameters to generate datasets of varying human subjects, as well as also use the pose parameters themselves as input to the network in one of the experiments. Since all the humans are in one-to-one correspondence, we can use their correspondences to compute the ground truth maps during training, while evaluating on different triangulations during evaluation.

\paragraph{Learning to re-pose humans.}
\begin{figure}
    \centering
    \includegraphics[width=\columnwidth]{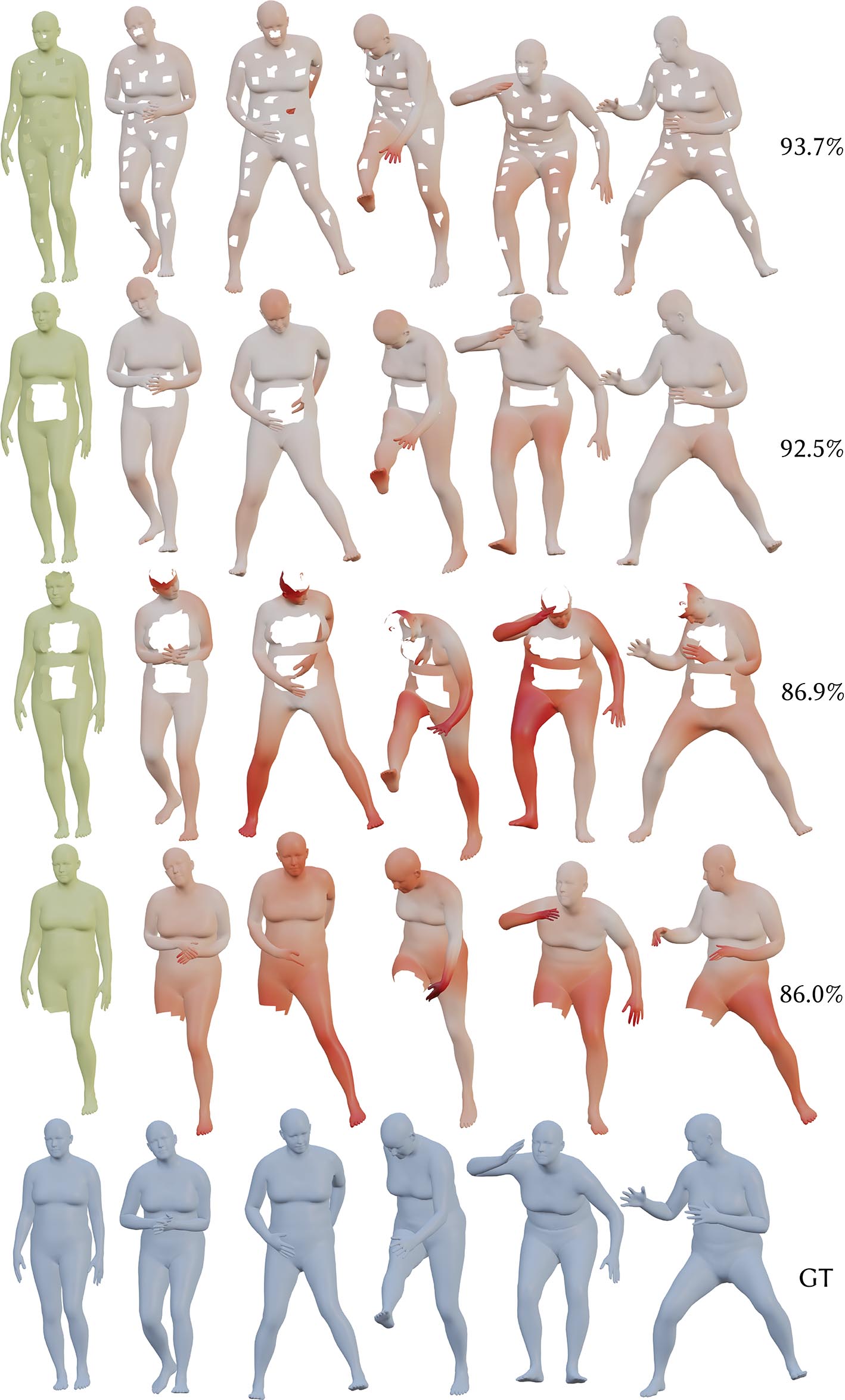}
    \vspace{-10pt}
    \caption{\textbf{Registering partial humans.} A network trained to register full SMPL~\cite{Bogo:ECCV:2016} models to one another manages predicts correct deformations (red) of source models (green) which have parts of them removed. Ground-truth shown in blue.}
     \vspace{-10pt}
    \label{fig:partial}
\end{figure}
By training the network to correctly re-pose  humans conditioned on pose parameters, we can enable it to control humanoid models. 
We constructed a dataset of $100$K random STAR meshes by sampling the parameter spaces. Each training sample consists of a human with arbitrary shape parameters in an arbitrary initial pose, along with arbitrary target pose parameters, describing the pose the human is to be re-posed to. We supply the network with a PointNet encoding of the initial pose and shape of the human along with pose parameters of the target pose, and optimize the loss $\totloss$ w.r.t. the ground truth re-posed human.
As Table \ref{tab:quantitative_results} shows, our network is able to map humans into the desired poses with  high accuracy. Results of a few examples are shown in Figure \ref{fig:pose_params}, with the source mesh shown in green. We (red) are indistinguishable from the ground truth pose (blue), as can be seen by the red color denoting error, with a maximum error below $0.05$. Note this is achieved with the network having only a PointNet encoding of  $1024$ sample-points from the input human, from which it needs to infer both correct shape as well as correct initial pose in order to successfully re-pose it.

With the network fully trained to deform any STAR model w.r.t. pose parameters, we apply it to Big Buck Bunny, a mesh with significantly different features to the training set, and a completely different triangulation. Results are shown in Figure \ref{fig:bunny}, showing the network truly disentangled pose from shape, and can apply the poses to a model which lies far outside of the training set's shape space.  Note the preservation of the details of the ears, even though they are a feature that is not present in any training sample of the humans. Likewise, note the natural deformation of the chin at the poses in which Big Buck is looking down.  

In terms of triangulation-agnosticism, we note that not only is Big Buck differently-triangulated to the training set, but in this case the network was trained solely on the one specific triangulation which all STAR meshes share, and it is due  to the framework's complete agnosticism that it did not overfit to that triangulation and could be applied to other meshes. 

On the other hand, training on STAR alone is insufficient for us in order to be able to generalize to any humanoid -- our method is highly-sensitive to the encoding of the input shape.
As a result,  the network is limited to meshes with similar profiles to the humans, and fails when applied to, e.g., the Armadillo. In order for the network to generalize to such case, it would need to be trained on humanoids with a wider range of profiles.

\paragraph{Full and partial registration.} 

We explore the sensitivity of our framework to changes in the triangulation between train and test data, through a synthetic partial-registration experiment. First, we generate a dataset of SMPL~\cite{Bogo:ECCV:2016} humans, by choosing a random identity, and $32$ different poses of it, and repeat this for multiple identities until we aggregate $100$K samples. We then choose pairs of the same identity in different poses, and train the network by feeding it a PointNet encoding of the two meshes and optimize $\totloss$. After the  network is fully trained, it can deform a given SMPL model in one pose and register it to another pose. Results are shown in Figure \ref{fig:smpl}. We then modify SMPL meshes from the test set by removing triangles from them to get partial meshes, and then feed each partial mesh as a source, paired with a PointNet encoding of the target. Results are shown in Figure \ref{fig:partial}. The network produces plausible deformations, and is in general unaffected by the holes, even though it was solely trained on a sphere-topology triangulation of the full model. In many cases the network manages to accurately match the partial surface to the full one (we visualize the error w.r.t. the ground truth), and in others fails gracefully by producing an inaccurate but detail-preserving and plausible deformation (note the heads),  however there is clear degradation as a result of the removal of triangles.  As the predicted jacobian field is conditioned on the PointNet encoding, it seems that the latter is the weak link as during training it has learned to latch on to strong cues of the limbs and head, causing the prediction to be much more sensitive to the absence of a leg than to that of the abdomen.

\begin{figure}
    \centering
    \includegraphics[width=\columnwidth]{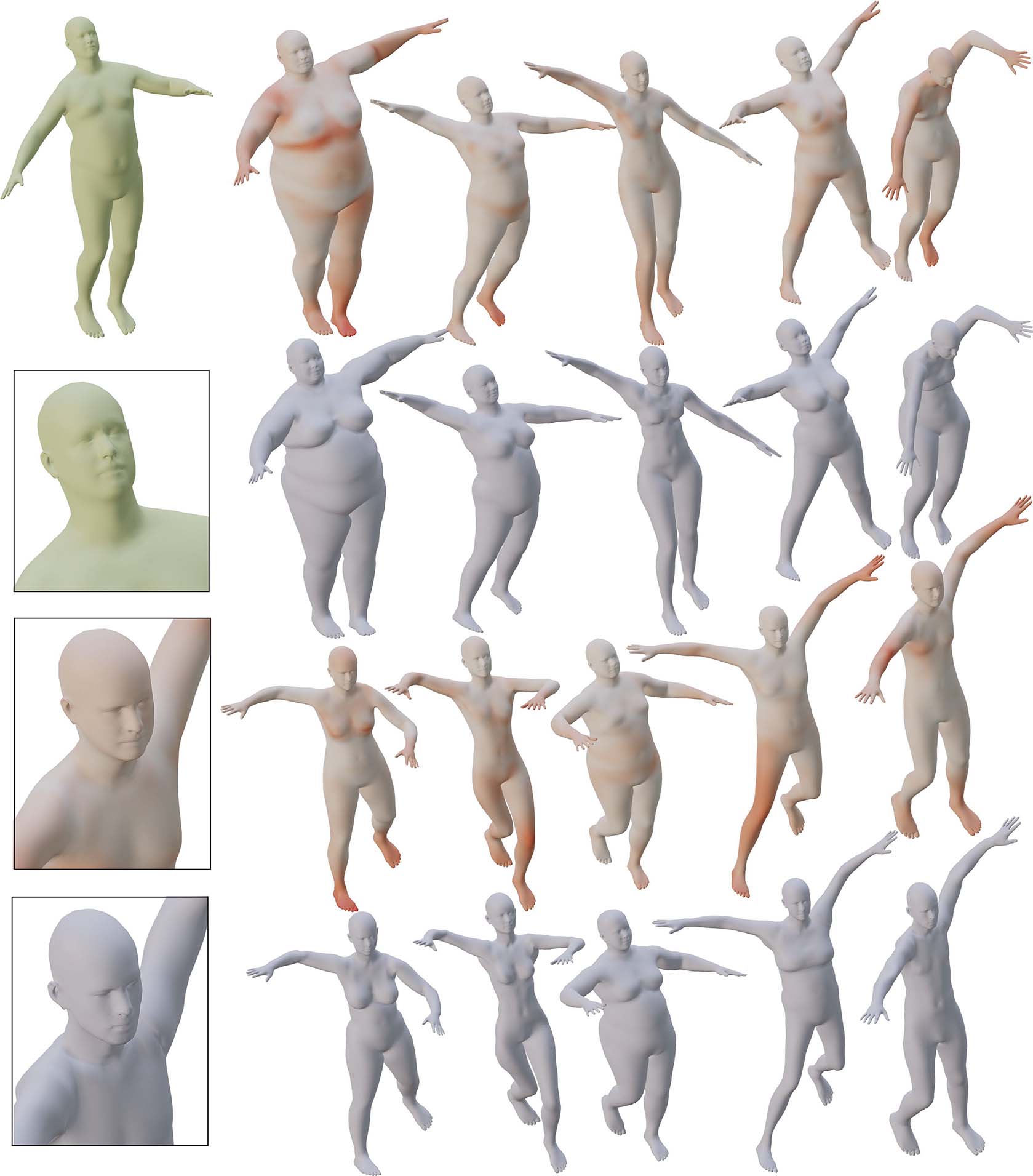}
    \caption{\textbf{Learning to morph humans into other humans.} We train our network to receive Pointnet~\cite{qi2017pointnet} encodings of two random humans in random poses from the STAR~\cite{STAR:2020} dataset, and map one into the other. Source mesh is shown in green, targets in blue and the deformations colored w.r.t. error (reddest being $0.1$).  The network correctly maps the humans, and in places where the pointnet encoding is not descriptive (e.g., the faces), defaults to an accurate detail-preserving deformation of the source.  }
    \label{fig:morph}
    \vspace{-10pt}
\end{figure}

\paragraph{Morphing humans.} In the next experiment, we train the network to morph one human into another, i.e., exactly match the surface of the target according to the ground truth correspondences between them (recall these correspondences are not given to the network as input). Each training sample consists of two random human identities in two random poses from the training set. We feed both of their PointNet encodings to the network, and train with the loss $\totloss$. Quantitative evaluations over the test set are shown in Table \ref{tab:quantitative_results}. We show  a few examples of a single source mapped to multiple different targets in Figure \ref{fig:morph}. Since the sampled points fed into the PointNet encoder are insufficient to accurately capture the facial features, our deformation network is not aware of the exact facial features of the target mesh and hence does not morph the faces. However, as can be seen in the zoom-ins in Figure \ref{fig:morph}, in lack of facial information, the intrinsic nature of the framework makes it default into a highly-detail preserving deformation of the head into the target pose, completely preserving the source mesh's face, while morphing the body correctly to the target pose. This could potentially be used by intentionally deleting features of the inputs during training, and forcing the network to ``hallucinate" the correct mapping, thereby effectively controlling which details of the original mesh are morphed into the target's, and which ones are preserved. 

\change{\paragraph{Discussion: relation to shape correspondence.}
The experiments in this section exhibit the ability of our network to infer deformation spaces of humans. This makes our framework a good candidate for performing shape-correspondence tasks (see~\cite{sahilliouglu2020recent} for a recent survey), which we set as an important future direction. Note that there is a subtle difference between reproducing plausible deformations and computing exact correspondences: for deformations, highly-accurate correspondences may still yield implausible deformations when considered as a mapping, e.g., a nose mapped to a cheek will result in a distorted face (alternatively, a plausible deformation cannot be translated to accurate correspondences necessarily, as closest-point matching can yield incorrect correspondences even for plausible deformations). Furthermore, note that we only receive as input $1024$ sampled points on the source/target and do not have a dense set of points over which to find dense correspondences.}

\subsection{Physical Simulation}
Data-driven methods are often considered in the context of physical simulation, e.g., for defining a simulation subspace as an alternative to performing direct optimization of the simulation's parameters with respect to the underlying physical model. Our framework can be used as a drop-in general tool in this setting,  without any modifications to the architecture to account for the underlying physical model.
\begin{figure}
    \centering
    \includegraphics[width=\columnwidth]{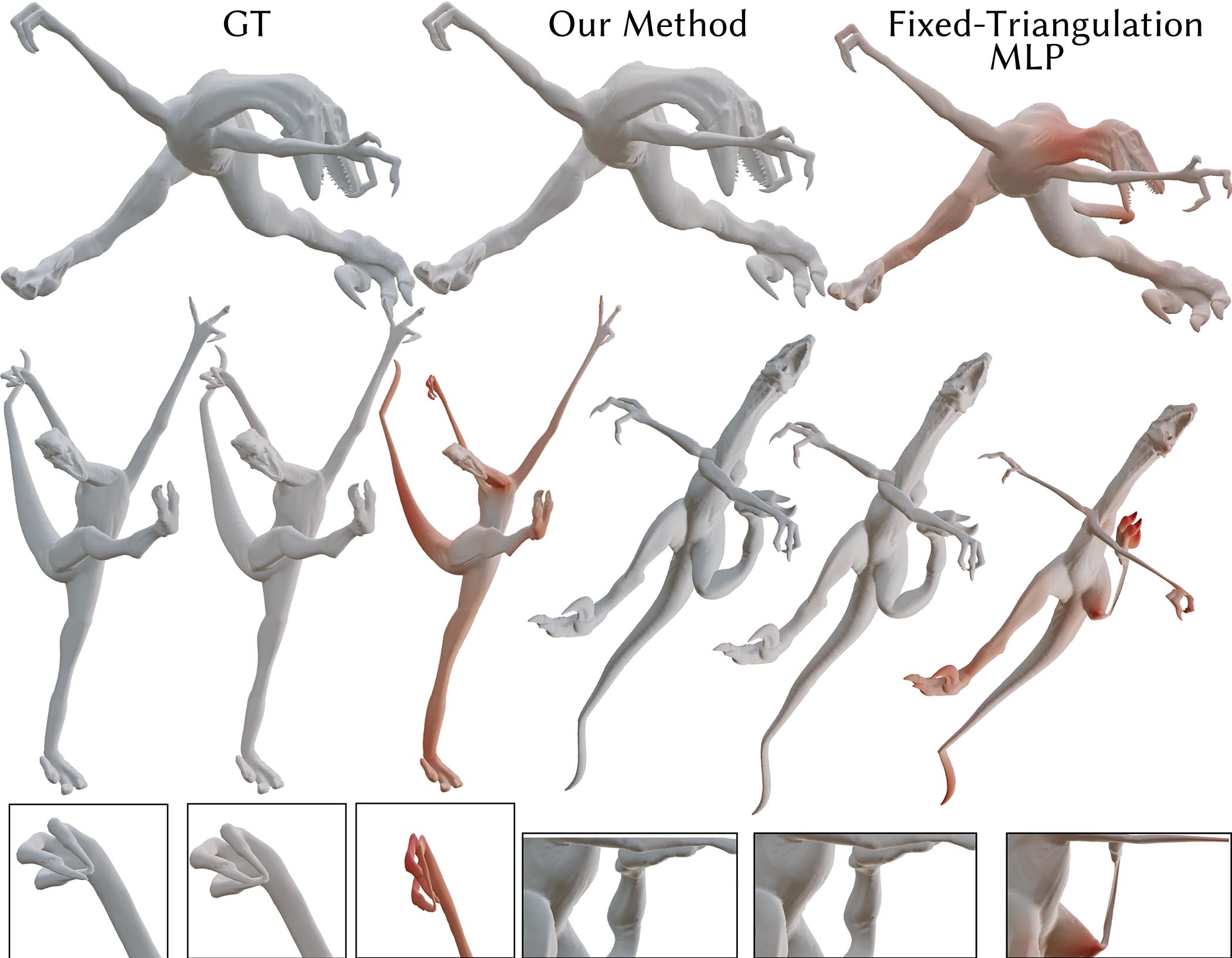}
    \caption{\textbf{Learning volumetric ARAP~\cite{sorkine2007arap} deformations}. We train our network to predict the boundary-surface deformation induced by volumetric ARAP deformations of a single model, conditioned  on the positions of predesignated constraint handles.  In each triplet we show, from left to right, the ground truth, us, and the result computed by a fixed-triangulation architecture, similar to \cite{tan2018meshvae}.}
    \label{fig:raptor}
\end{figure}

\paragraph{Learning volumetric ARAP~\cite{sorkine2007arap} deformations.} We generated a dataset of volumetric ARAP deformations of a model by tetrahedralizing the Raptor model into a tetrahedral mesh consisting of one million elements. We choose $6$ fixed handles within the limbs, head and tail of the raptor. Then, we generate $100$K training samples, by shifting the handles into random configurations, computing the corresponding ARAP deformation, and saving the boundary triangle mesh, while discarding the tetrahedra. The network is then given as input the 3D positions of the handles, and is trained  with the loss $\totloss$ w.r.t. the ground truth deformation. Quantitative results are reported in Table \ref{tab:quantitative_results}, showing high accuracy.  We show a few  results in Figure \ref{fig:raptor}. We are able to accurately capture various modes and behaviors of ARAP, such as bending and stretching. ARAP's optimization converges in 10 minutes; our feedforward and linear solve requires approximately a second, using Pytorch and Numpy without any further optimization. We also compare our results to the performance of a fixed-triangulation gradient domain method similar to \cite{tan2018meshvae}. We discuss it in Subsection \ref{sec:comparisons}.

\paragraph{Learning collision handling}
\begin{figure}
    \centering
    \includegraphics[width=\columnwidth]{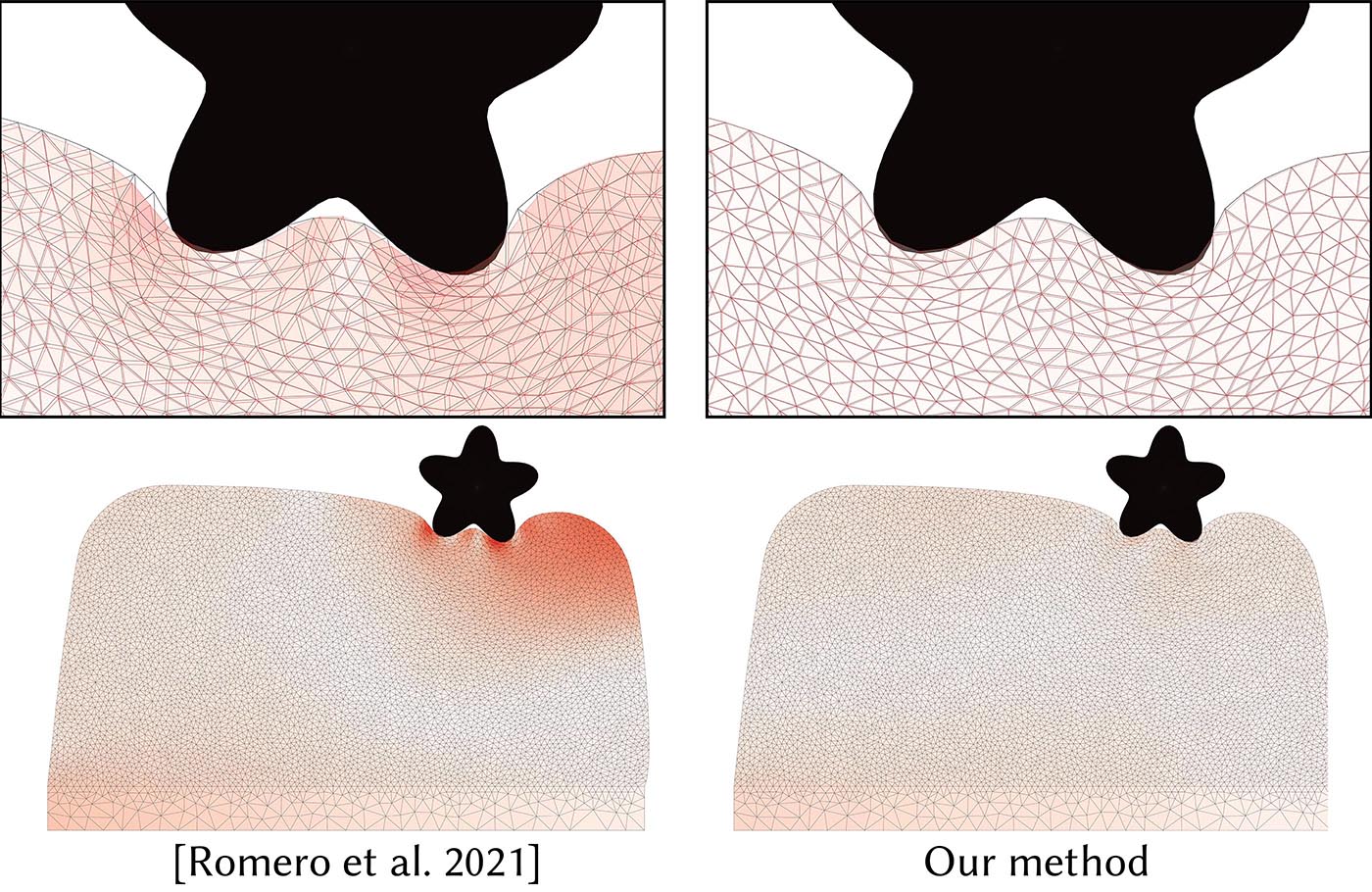}
    \caption{\textbf{Learning collision handling.} We compare our method to \cite{Romero:2021} on collision handling, using their data to train and test. Predictions are overlaid over the ground-truth in the zoom-ins. We achieve slightly higher accuracy, but at higher running times.}
    \label{fig:jelly}
\end{figure}
\cite{Romero:2021} propose an elegant method to train networks to produce non-linear offsets from a linear deformation model so as to account for non-linear behavior due to object collisions, by feeding the network the positions of the collider, along with the handles' positions. We train on their data with the same input as them and optimize for $\totloss$, without introducing any changes to our architecture. Results are shown in Figure \ref{fig:jelly}, with a numerical evaluation in Table \ref{tab:quantitative_results}. We achieve slightly more accurate results than Romero, showing our framework is highly versatile. \cite{Romero:2021} is however tailored to be used within simulation frameworks and thus achieve a much higher FPS rate than us. Note that in this case the problem is restricted to 2D, thereby trivializing our restriction operator $\proj_i$. In light of this, the experiment solely highlights the efficacy  of learning jacobians as a continuous smooth field, instead of a discrete prediction.

\subsection{Comparisons}
\label{sec:comparisons}
We now move on to comparing our method to other techniques for mapping and deformation. 
\begin{figure}
    \centering
    \includegraphics[width=\columnwidth]{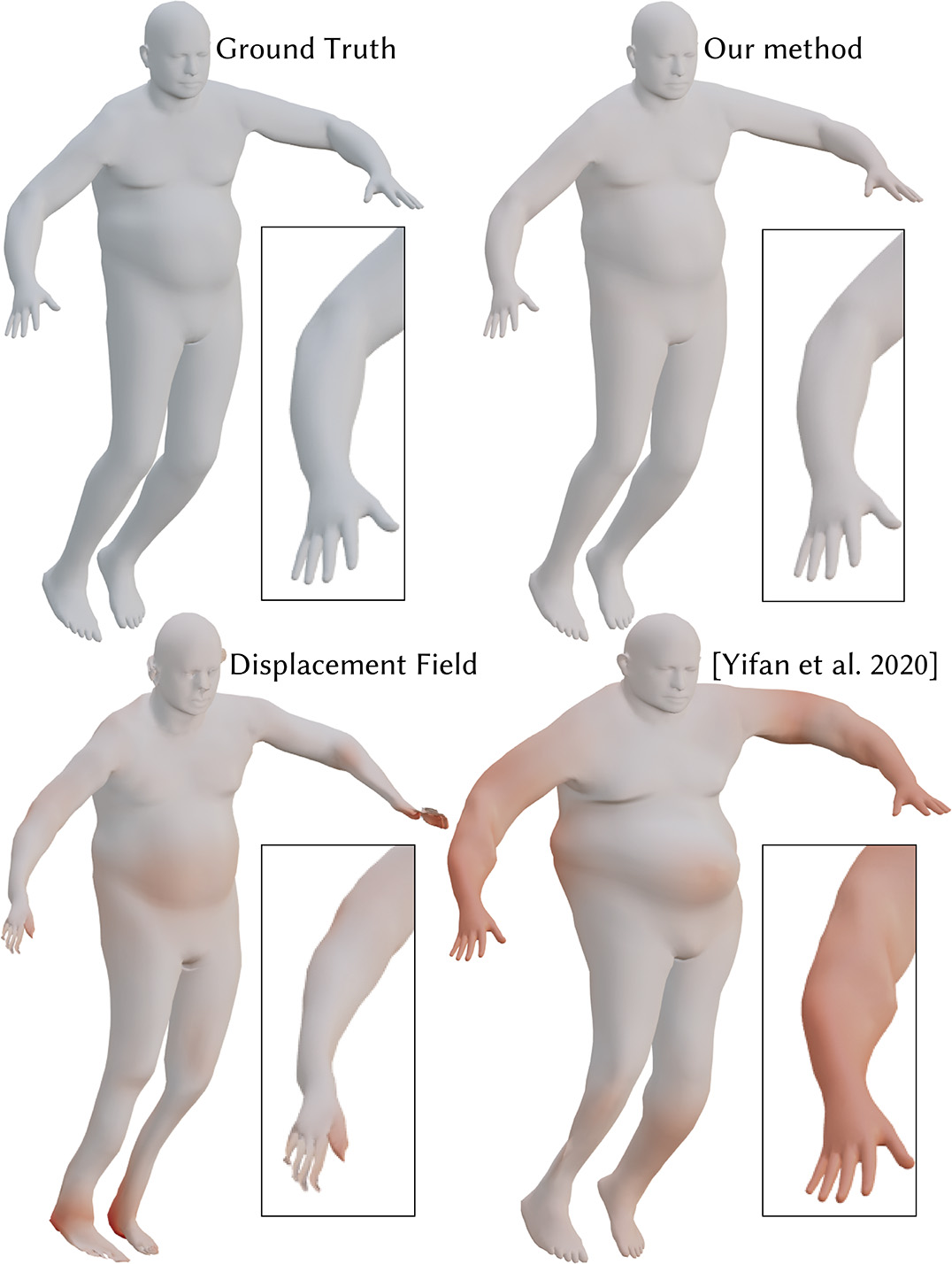}
    \caption{\textbf{Comparison to Neural Cages~\cite{wang2019neural} and direct prediction of a displacement field.} Our method accurately re-poses the source, while both the displacement field approach and Neural Cages distort the shape.}
    \label{fig:ncages}
\end{figure}

\paragraph{Neural Cages~\cite{wang2019neural}.} Similarly to us, rigging-based methods such as Neural Cages are triangulation-agnostic.  Cage-based deformations reduce the deformation space in a manner that excludes highly-oscillatory solutions, but still require a highly accurate prediction of the rig in order to reproduce an artifact-free deformation. 

We compare our method to the deformation module of Neural Cages, on a variant of the re-posing experiment from Section \ref{subsec:re-posing_and_registration}. To accommodate for the fitting of the cage required by \cite{wang2019neural}, we opt to have each source model in the T-pose. Note that in the original paper's humanoid experiments, Yifan et al. do not train the network to fit the cage to the source mesh. Instead, they use given manual annotations to obtain a valid cage at training and at inference time. We follow this setup as well.

To create a good cage  for each identity, we first construct a high-quality, enveloping cage over one source in the T-pose. We then use the ground truth vertex correspondences between the STAR models to transfer the MVC weights from the original source to the new source, thereby providing the global optimum to the same fitting described in the original paper. We further confirm that the cages generated in this manner can yield correct deformations matching the source to the target, by overfitting to one example. Note that this gives a significant advantage to Neural Cages since it prevents it from being affected by errors in the cage predictions, which would happen without using the ground truth STAR correspondences. Both ours and their method are then trained over the training set, which consists of pairs of an arbitrary human identity in the T-pose, along with star parameters which describe the pose to deform the human into.   We  train both methods using the ground truth correspondences between the different STAR models as losses.  

As Figure \ref{fig:ncages} shows, their network struggles to accurately deform  the  cage correctly for each source, as the cage's deformation needs to be highly accurate itself to not introduce artifacts, with slight variations in it resulting in artifacts in the deformation. This is numerically validated in Table \ref{tab:quantitative_results}. We do note, however, that there exist scenarios in which our intrinsic shape-awareness also has disadvantages: their network, trained on humans, could be applicable to drastically different characters, such as a robot, since  the cage warps ambient space and the surfaces embedded in it, without being affected by their intrinsic properties.  Our method, on the other hand, would fail on such a model, as intrinsically it is completely different to  the training set. 

\begin{table}[t]
\centering
\small
\resizebox{\linewidth}{!}{
\begin{tabular}{lccccc}
\toprule
    & L2-V  $\downarrow$ & L2-J  $\downarrow$ & L2-N $\downarrow$ & Hz $\uparrow$  & Fig. \\
    % \textit{\ \ \ Romero []} & - & - & -  \\ 
    \\
    \textbf{Applications} (all ours) & & & &\\
    \midrule
    % Learning UV
    \textbf{UV parameterization} & & & & \\
    \textit{\ \ \  Thingi10K~\cite{zhou2016thingi10k}} & 4.66 & 3.68 & - & 142 & \ref{fig:uv_all_in_one} \\
    \textit{\ \ \  UV-generalization} & 3.49 & 2.59 & - & 84 & \ref{fig:uv_all_in_one} \\

    % star complete random
    % star random to random via handles
    \textbf{Humans} & & & & \\
    \textit{\ \ \  Re-posing STAR~\cite{STAR:2020}} & 1.84 & 1.55 & 4.4 & 93 &  \ref{fig:pose_params} \\
    \textit{\ \ \  Morphing STAR~\cite{STAR:2020}} & 2.00 & 2.80 & 8.2 &  85  & \ref{fig:morph}\\
    \textit{\ \ \ Registration SMPL~\cite{Bogo:ECCV:2016}} & 3.24 & 3.01 & 9.9 & 85 & \ref{fig:smpl} \\
    \\
    \textbf{Comparisons} & & & & \\
    \midrule
    \textbf{ARAP~\cite{sorkine2007arap}} & & & & \\
    \textit{\ \ \ Global MLP baseline} & 3.66 & 1.82 & 13.6 & - & \ref{fig:raptor}\\
    \textit{\ \ \ Ours} & \textbf{0.64} & \textbf{0.43} &  \textbf{2.7} & - & \ref{fig:raptor}  \\
    \textbf{Collision Handling} & & & & \\
    \textit{\ \ \ ~\cite{Romero:2021}}& 0.424 & - & - &  \textbf{980} & \ref{fig:jelly} \\
    \textit{\ \ \ Ours} &  \textbf{0.398} &  \textbf{0.17} & - & 109 & \ref{fig:jelly} \\
    
    % Star to 32
    \textbf{Re-posing} & & & & \\
    \textit{\ \ \ Neural Cages~\cite{wang2019neural}} & 3.87 & 1.56 & 9.6 & - & \ref{fig:ncages} \\
   %| L2 | 0.042965410679744256 | 0.013457937485268832
%| J | 0.5647949766369886 | 0.11447658740189878
%| N | 0.4611665924239443 | 0.06488413246179454
    \textit{\ \ \ Displacement Field} & 4.21 & 5.62 & 11.1 & - & \ref{fig:ncages} \\
   
   \textit{\ \ \ Ours} &  \textbf{0.84} &  \textbf{0.83} & 2.4 & 87 &  \ref{fig:ncages}\\
\bottomrule
\end{tabular}
}
\caption{
{
\textbf{Quantitative comparisons on various tasks and datasets.}  \\ 
We report the L2 distance between prediction and ground-truth, summed over the mesh's vertices, after normalizing to a unit sphere and scaling by $10^2$ (\textbf{L2-V}), summer over the Jacobian matrices, scaled by $10$ (\textbf{L2-J}), and the average angular error on the face normals in degrees (\textbf{L2-N}). We also show the number of feed-forward inferences (\textbf{Hz}) per second using a single Nvidia V100 and a batch size of $1$. All results are reported on unseen data.
}
}
\label{tab:quantitative_results}
\end{table}

\begin{figure*}
    \centering
    \includegraphics[width=\textwidth]{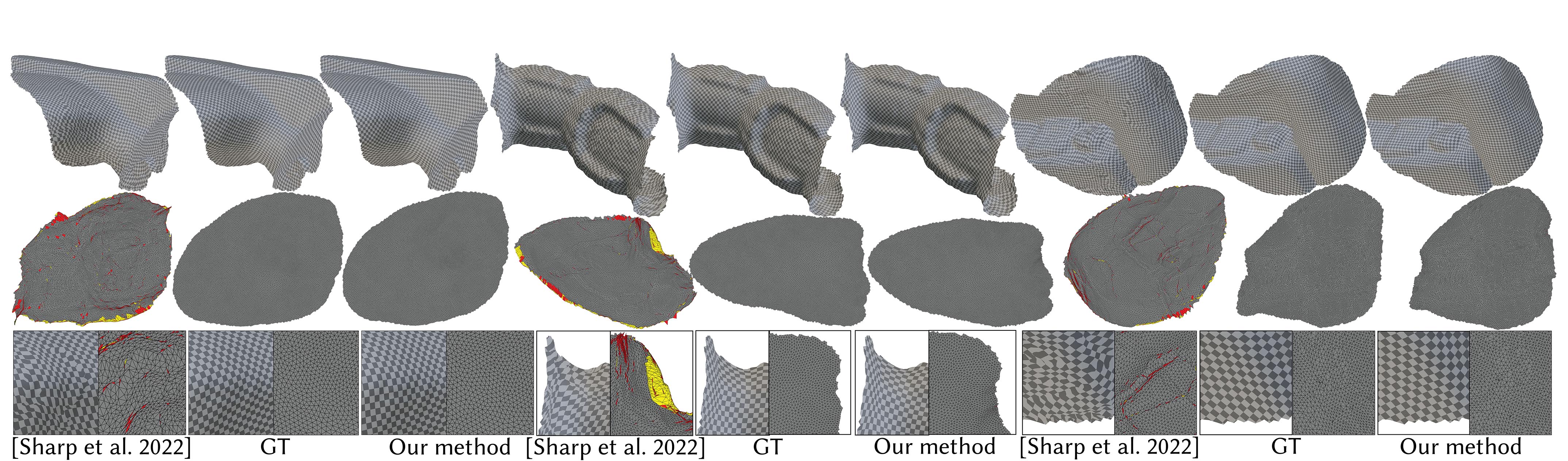}
    \caption{\change{\textbf{Comparison to DiffusionNet~\cite{sharp2022diffusionnet}.} We train DiffusionNet on the UV-learning experiment from Figure \ref{fig:uv_all_in_one}, and show its predictions on the test set, along with our network's output and the ground truth produced by SLIM~\cite{Rabinovich:SLIM:2017}. We highlight inverted triangles in yellow and triangles with distortion higher than $3$ in red. Our output is consistent with SLIM's, and exhibits no high distortion nor inverted triangles.}}
    \label{fig:diffnet}
\end{figure*}
\paragraph{Displacement field.} \change{Instead of predicting a field of jacobians, our MLP could directly output displacements for each point in space, similarly to~\cite{jiang2020shapeflow,huang2020meshode}. This approach is commonly used in mesh-agnostic settings. The limitation of this method lies in it not being shape-aware. This can be observed by noting that for a predicted displacement field, the displacement of a given point is the same, regardless of the shape that is mapped. Thus the prediction itself needs to account for the details of the underlying geometry, resulting in the network struggling to make accurate predictions. In contrast, a single, predicted jacobian field would produce \emph{different}, detail-preserving displacements for the same point in space, for different shapes, as each shape admits a different Laplacian, thereby lifting the burden of the detail preservation from the network. These assertions are verified through qualitative and quantitative comparisons in Figure \ref{fig:ncages} and Table \ref{tab:quantitative_results}, respectively.}\\

\emph{Fixed-triangulation gradient-domain methods.} Instead of using our MLP, which receives one triangle and outputs one jacobian at a time,  previous works that operate in the gradient domain such as Tan et al.~\cite{tan2018meshvae} and~\cite{gaovcgan2018} use a global MLP which outputs a single, global prediction of a tensor, in our case of size $\abs{T}\times3\times2$, consisting of stacked jacobians. These  are then assigned in an arbitrary, consistent order to the triangles. As discussed before, such a construction is fundamentally not applicable to all the experiments shown here, which consist of datasets with more than one triangulation, placing our method and theirs in different categories. 

However, putting aside triangulation-agnosticism, we wish to argue for the effectiveness of learning a jacobian \emph{field}, instead of making a prediction of a global tensor. As discussed, the gradient of most maps considered in geometry-processing applications is, relatively, a gradually-varying, low-frequency signal over the domain,  ideal for regression and learning. However, using a global MLP completely discards all spatial knowledge. This is a burden on the MLP, causing it to expand its capacity on re-inferring spatial relationships through the fully-connected architecture and the training losses. To exhibit this, we replace our architecture with a global MLP predicting a tensor of jacobians and compare it to ours on learning the Raptor's ARAP deformation space, discussed above. We report the quantitative comparison in Table \ref{tab:quantitative_results}, and visually compare the global MLP's produced deformations to ours (Figure \ref{fig:raptor}), showing in each example from left to right the ground truth, our network's prediction, and the global MLP's. As the raptor consists of 70K triangles, the global MLP lacks the capacity to make an accurate prediction. In contrast, our method leverages the smooth spatial behavior of the jacobians and produces accurate results. Furthermore, our network does not need to increase its capacity with respect to the density of the triangulation, rather only with respect to the underlying granularity of details on the source mesh. 

\paragraph{DiffusionNet~\cite{sharp2022diffusionnet}.} \change{Techniques such as DiffusionNet define convolution-like operators to process signals defined over the surface. Similarly to us, they are also triangulation-agnostic and can be applied to arbitrary meshes, which makes them a possible candidate for, e.g., the UV mapping experiment; we compare our performance to theirs on this experiment, by training their method on the same training set as ours and evaluating on the same test sets. The main difference between the two methods lies in ours making a local, per-point prediction through an MLP, along with leveraging Poisson's equation to predict a smooth signal in the gradient domain which guides the mapping; \cite{sharp2022diffusionnet} produce a direct global prediction of the output through diffusion operators, and without using Poisson's equation, which ``integrates'' predicted gradients. Hence, they struggle to make accurate predictions on the delicate UV-mapping task. As shown in Table \ref{tab:uv}, their predictions exhibit significantly higher distortion, as well as significantly more inverted triangles. We show qualitative comparisons on a few examples, exhibiting these artifacts in Figure \ref{fig:diffnet}. We highlight inverted elements in yellow and triangles with distortion higher than $3$ in red. Our method's output closely matches the ground truth, exhibits barely any high distortion, and no inverted triangles, while \cite{sharp2022diffusionnet} produces a highly-oscillatory output, which does not match the ground truth and exhibits high distortion and inverted elements.}   

\change{We note that using DiffusionNet to produce input feature-vectors to our MLP instead of the Wave Kernel Signature, or swapping it in place of the PointNet encoder stand as appealing future directions that may leverage the benefits of both approaches.}

\begin{figure*}
    \centering
    \includegraphics[width=\textwidth]{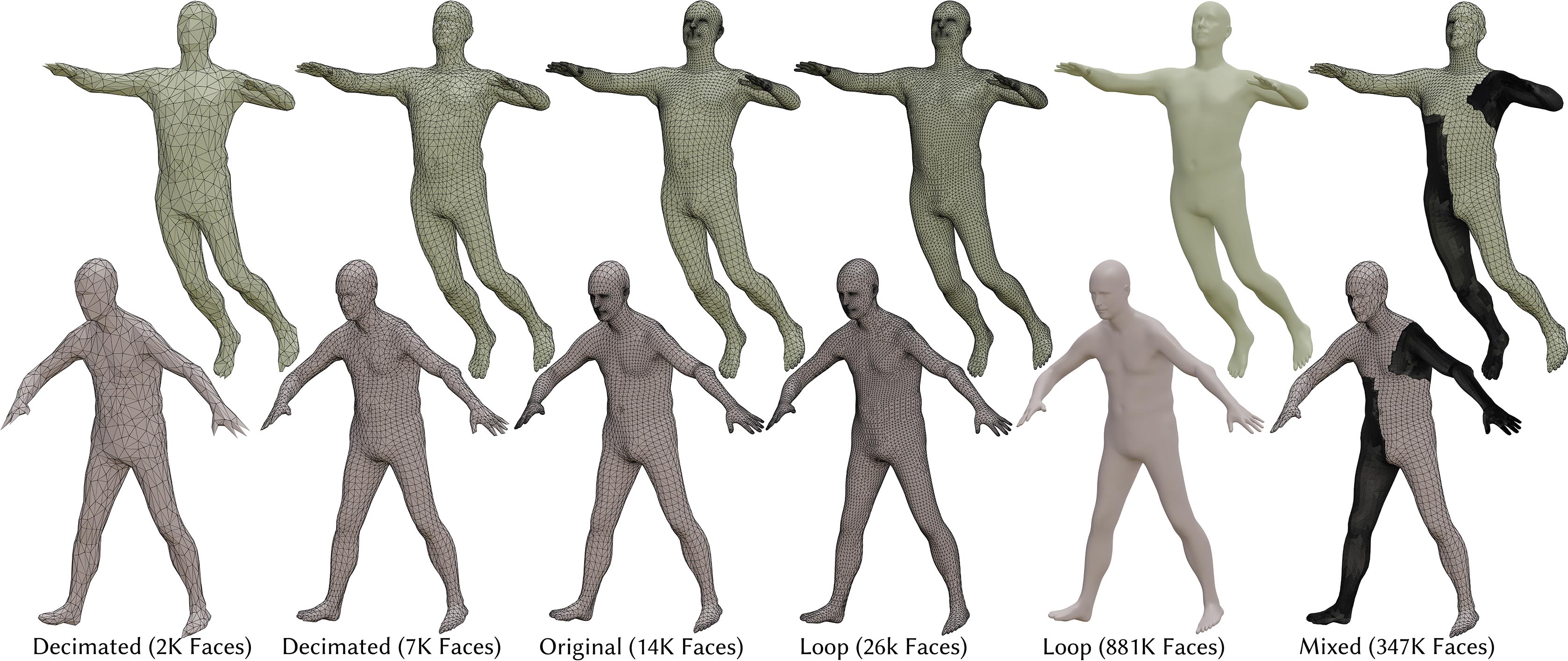}
    \caption{\change{\textbf{Triangulation agnosticism of the framework.} Our network, trained for the re-posing experiment shown in Figure~\ref{fig:pose_params}, produces the same predictions (bottom row) for different triangulations of the source mesh (top row), even though it was trained on the single, canonical triangulation of the STAR~\cite{STAR:2020} dataset (labeled ``Original'' here).}  }
    \label{fig:diff_tri}
\end{figure*}

\subsection{Triangulation Agnosticism}
\change{While the losses we use during training are derived from the specific triangulation of the meshes, the MLP itself is unaware of anything more than a single triangle during inference, thereby making the network completely agnostic to the triangulation. We validate this claim by applying the network trained for the re-posing experiment, Section \ref{subsec:re-posing_and_registration}, on different remeshings of the same surface, from decimated ones to ones achieved via Loop subdivision, as well as a ``mixed-technique'' model with varying triangulation density. Results are shown in Figure \ref{fig:diff_tri}. As expected, the network reproduces approximately the same deformation for all the different meshes.  }

\change{The ability of the network to empirically be almost invariant to the various triangulation can be understood intuitively from a few simple observations: 1) The PointNet encoder used in this experiment to encode the shape is, broadly-speaking, invariant to triangulations since the points fed to it are randomly sampled from the surface; 2) Given the encoding of the shape, the MLP predicts a smooth, gradually-varying field of jacobians, and hence, while different triangulations sample this continuous field in different locations, for dense-enough samples the sampling will converge; 3) the differential operators of the meshes are known (e.g., through Cea's Lemma) to converge as the triangulation is refined, hence the differential operators of the different triangulations produce similar results when applied to the jacobian field.}

\change{Lastly, we emphasize the difference between our framework's triangulation-\emph{agnosticism} (lack of knowledge of the triangulation prevents the network from fitting to one triangulation) and a hypothetical complete \emph{invariance} to the triangulation (network is guaranteed to produce exactly same output for any triangulation): our method is FEM-based, and modifying the triangulation modifies the space of piecewise-linear maps, i.e., the space of possible outputs. Therefore, triangulation-invariance is not well-defined for this case.}

\section{Conclusion}
 The experiments shown in this paper validate that our framework for predicting piecewise-linear mappings of meshes produces highly-accurate and plausible maps on heterogeneous collections of meshes with varying triangulations. They also serve to exhibit the framework's high versatility, as it did not require any modifications for its application to any of the broad range of scenarios considered herein.

\emph{Future work.} A few important research directions lie ahead. First, in this paper we have focused on the novel deformation module, and hence opted to use a rather-naive approach to encoding input shapes via a PointNet encoder, which we do not consider inherent to our framework.  Exploring encoders is an important next step, e.g., using intrinsic ones such as \cite{sharp2022diffusionnet}, or image-based ones. We are also highly-interested in exploring unsupervised-learning scenarios next, with the immediate candidates being optimizing various distortion and quality measures, or coupling our framework with a visual loss from a neural-renderer. 

many of the applications exhibited here are worth revisiting more deeply, tailoring our framework more specifically to them. There are also additional immediate practical applications, e.g., training to map smoothed shapes onto their original geometry, thereby learning a latent space of details which could then be applied to other meshes. 

Additionally, one technical consequence of our framework is that it can be used as a differentiable layer, enabling various losses defined in terms of map-predictions. For instance, it is highly enticing to define a differentiable loss for a source mesh, which measures how well does our network map it to a given target, and then optimize the source mesh to reduce that loss and better fit the target. This could be useful for, e.g., optimizing the mesh to accommodate for its desired deformation space. Lastly, we note our approach is directly extendable to tetrahedral-meshes and 2D meshes, in which case the restriction operator is not needed.
\paragraph{}{  Limitations.} While our network is highly effective in computing mappings of triangulated meshes, it relies heavily on  discrete differential geometry operators, namely the use of the gradient and Laplacian operators. Hence, in cases where ones are not readily available, such as on point clouds or non-manifold meshes, our network may not be applicable. This may be resolved by using an estimated manifold proxy to one of those operators. \change{Additionally, as the MLP receives positions and normals in cartesian coordinates, it is not invariant to rigid motions nor scaling of the input. This can be resolved by either choosing canonical orientations (e.g., via PCA), augmenting the training data, or using solely intrinsic descriptors and encoders.}

Furthermore, we did not explore handling multiple connected components. Our framework is readily applicable to that scenario, by having the MLP predict in addition the matrix $\pred_i$ a translation vector, and then averaging these translations per each connected component to yield a global translation of each one of them. However, as the pointnet encoder is unaware of topology, the encoding it yields does not enable the network to discern topological information, e.g., the clothes of a human from the human itself, and the training process will fail. This reiterates the necessity of devising an intrinsic encoder of shapes. 

Lastly, we note that the current formulation enables us to only produce continuous maps. In case a discontinuous one is desired, e.g., in order to introduce cuts to a UV parameterization, additional means such as masking are needed. \change{We also note that the MLP produces  a \emph{continuous} jacobian field, and hence, theoretically, cannot strictly represent a sharp discontinuity in derivatives (e.g., when a flat plane needs to bend by 90 degrees); however, it can predict a smooth approximation of it, and since discrete triangular meshes admit deformation spaces which model sharp discontinuities even when the underlying predicted continuous field (sampled discretely on the mesh) is smooth, we did not experience any such failure cases in practice.}

\bibliographystyle{ACM-Reference-Format}
\bibliography{biblio}


\begin{thebibliography}{79}

%%% the Web address.

\ifx \showCODEN    \undefined \def \showCODEN     #1{\unskip}     \fi
\ifx \showDOI      \undefined \def \showDOI       #1{#1}\fi
\ifx \showISBNx    \undefined \def \showISBNx     #1{\unskip}     \fi
\ifx \showISBNxiii \undefined \def \showISBNxiii  #1{\unskip}     \fi
\ifx \showISSN     \undefined \def \showISSN      #1{\unskip}     \fi
\ifx \showLCCN     \undefined \def \showLCCN      #1{\unskip}     \fi
\ifx \shownote     \undefined \def \shownote      #1{#1}          \fi
\ifx \showarticletitle \undefined \def \showarticletitle #1{#1}   \fi
\ifx \showURL      \undefined \def \showURL       {\relax}        \fi
% invisible to TeX
\providecommand\bibfield[2]{#2}
\providecommand\bibinfo[2]{#2}
\providecommand\natexlab[1]{#1}
\providecommand\showeprint[2][]{arXiv:#2}

\bibitem[Aigerman and Lipman(2013)]%
        {aigerman2013injective}
\bibfield{author}{\bibinfo{person}{Noam Aigerman} {and} \bibinfo{person}{Yaron
  Lipman}.} \bibinfo{year}{2013}\natexlab{}.
\newblock \showarticletitle{Injective and bounded distortion mappings in 3D}.
\newblock \bibinfo{journal}{\emph{ACM Transactions on Graphics (TOG)}}
  \bibinfo{volume}{32}, \bibinfo{number}{4} (\bibinfo{year}{2013}),
  \bibinfo{pages}{1--14}.
\newblock


\bibitem[Anguelov et~al\mbox{.}(2005)]%
        {anguelov2005scape}
\bibfield{author}{\bibinfo{person}{Dragomir Anguelov}, \bibinfo{person}{Praveen
  Srinivasan}, \bibinfo{person}{Daphne Koller}, \bibinfo{person}{Sebastian
  Thrun}, \bibinfo{person}{Jim Rodgers}, {and} \bibinfo{person}{James Davis}.}
  \bibinfo{year}{2005}\natexlab{}.
\newblock \showarticletitle{{SCAPE}: Shape Completion and Animation of People}.
  In \bibinfo{booktitle}{\emph{SIGGRAPH}}.
\newblock


\bibitem[Aubry et~al\mbox{.}(2011)]%
        {aubry2011wave}
\bibfield{author}{\bibinfo{person}{Mathieu Aubry}, \bibinfo{person}{Ulrich
  Schlickewei}, {and} \bibinfo{person}{Daniel Cremers}.}
  \bibinfo{year}{2011}\natexlab{}.
\newblock \showarticletitle{The wave kernel signature: A quantum mechanical
  approach to shape analysis}. In \bibinfo{booktitle}{\emph{2011 IEEE
  international conference on computer vision workshops (ICCV workshops)}}.
  IEEE, \bibinfo{pages}{1626--1633}.
\newblock


\bibitem[Bailey et~al\mbox{.}(2020)]%
        {bailey2020fast}
\bibfield{author}{\bibinfo{person}{Stephen~W Bailey}, \bibinfo{person}{Dalton
  Omens}, \bibinfo{person}{Paul Dilorenzo}, {and} \bibinfo{person}{James~F
  O'Brien}.} \bibinfo{year}{2020}\natexlab{}.
\newblock \showarticletitle{Fast and deep facial deformations}.
\newblock \bibinfo{journal}{\emph{ACM Transactions on Graphics (TOG)}}
  \bibinfo{volume}{39}, \bibinfo{number}{4} (\bibinfo{year}{2020}),
  \bibinfo{pages}{94--1}.
\newblock


\bibitem[Bailey et~al\mbox{.}(2018)]%
        {Bailey:2018:FDD}
\bibfield{author}{\bibinfo{person}{Stephen~W. Bailey}, \bibinfo{person}{Dave
  Otte}, \bibinfo{person}{Paul Dilorenzo}, {and} \bibinfo{person}{James~F.
  O'Brien}.} \bibinfo{year}{2018}\natexlab{}.
\newblock \showarticletitle{Fast and Deep Deformation Approximations}.
\newblock \bibinfo{journal}{\emph{ACM Transactions on Graphics}}
  \bibinfo{volume}{37}, \bibinfo{number}{4} (\bibinfo{date}{Aug.}
  \bibinfo{year}{2018}), \bibinfo{pages}{119:1--12}.
\newblock
\urldef\tempurl%
\url{https://doi.org/10.1145/3197517.3201300}
\showDOI{\tempurl}
\newblock
\shownote{Presented at SIGGRAPH 2018, Los Angeles}.


\bibitem[Baran and Popovi\'{c}(2007)]%
        {Baran:rigging:2007}
\bibfield{author}{\bibinfo{person}{Ilya Baran} {and} \bibinfo{person}{Jovan
  Popovi\'{c}}.} \bibinfo{year}{2007}\natexlab{}.
\newblock \showarticletitle{Automatic Rigging and Animation of 3D Characters}.
\newblock \bibinfo{journal}{\emph{ACM Trans. Graph.}} \bibinfo{volume}{26},
  \bibinfo{number}{3} (\bibinfo{date}{jul} \bibinfo{year}{2007}),
  \bibinfo{pages}{72–es}.
\newblock
\showISSN{0730-0301}
\urldef\tempurl%
\url{https://doi.org/10.1145/1276377.1276467}
\showDOI{\tempurl}


\bibitem[Bednarik et~al\mbox{.}(2020)]%
        {bednarik2020shape}
\bibfield{author}{\bibinfo{person}{Jan Bednarik}, \bibinfo{person}{Shaifali
  Parashar}, \bibinfo{person}{Erhan Gundogdu}, \bibinfo{person}{Mathieu
  Salzmann}, {and} \bibinfo{person}{Pascal Fua}.}
  \bibinfo{year}{2020}\natexlab{}.
\newblock \showarticletitle{Shape reconstruction by learning differentiable
  surface representations}. In \bibinfo{booktitle}{\emph{Proceedings of the
  IEEE/CVF Conference on Computer Vision and Pattern Recognition}}.
  \bibinfo{pages}{4716--4725}.
\newblock


\bibitem[Bogo et~al\mbox{.}(2016)]%
        {Bogo:ECCV:2016}
\bibfield{author}{\bibinfo{person}{Federica Bogo}, \bibinfo{person}{Angjoo
  Kanazawa}, \bibinfo{person}{Christoph Lassner}, \bibinfo{person}{Peter
  Gehler}, \bibinfo{person}{Javier Romero}, {and} \bibinfo{person}{Michael~J.
  Black}.} \bibinfo{year}{2016}\natexlab{}.
\newblock \showarticletitle{Keep it {SMPL}: Automatic Estimation of {3D} Human
  Pose and Shape from a Single Image}. In \bibinfo{booktitle}{\emph{Computer
  Vision -- ECCV 2016}} \emph{(\bibinfo{series}{Lecture Notes in Computer
  Science})}. \bibinfo{publisher}{Springer International Publishing}.
\newblock


\bibitem[Bogo et~al\mbox{.}(2014)]%
        {bogo2014faust}
\bibfield{author}{\bibinfo{person}{Federica Bogo}, \bibinfo{person}{Javier
  Romero}, \bibinfo{person}{Matthew Loper}, {and} \bibinfo{person}{Michael~J.
  Black}.} \bibinfo{year}{2014}\natexlab{}.
\newblock \showarticletitle{{FAUST}: Dataset and evaluation for {3D} mesh
  registration}. In \bibinfo{booktitle}{\emph{CVPR}}.
\newblock


\bibitem[Bouaziz et~al\mbox{.}(2016)]%
        {Sofien:2016}
\bibfield{author}{\bibinfo{person}{Sofien Bouaziz}, \bibinfo{person}{Andrea
  Tagliasacchi}, \bibinfo{person}{Hao Li}, {and} \bibinfo{person}{Mark Pauly}.}
  \bibinfo{year}{2016}\natexlab{}.
\newblock \showarticletitle{Modern Techniques and Applications for Real-Time
  Non-Rigid Registration}. In \bibinfo{booktitle}{\emph{SIGGRAPH ASIA 2016
  Courses}} (Macau) \emph{(\bibinfo{series}{SA '16})}.
  \bibinfo{publisher}{Association for Computing Machinery},
  \bibinfo{address}{New York, NY, USA}, Article \bibinfo{articleno}{11},
  \bibinfo{numpages}{25}~pages.
\newblock
\showISBNx{9781450345385}
\urldef\tempurl%
\url{https://doi.org/10.1145/2988458.2988490}
\showDOI{\tempurl}


\bibitem[Du et~al\mbox{.}(2020)]%
        {du2020lifting}
\bibfield{author}{\bibinfo{person}{Xingyi Du}, \bibinfo{person}{Noam Aigerman},
  \bibinfo{person}{Qingnan Zhou}, \bibinfo{person}{Shahar~Z Kovalsky},
  \bibinfo{person}{Yajie Yan}, \bibinfo{person}{Danny~M Kaufman}, {and}
  \bibinfo{person}{Tao Ju}.} \bibinfo{year}{2020}\natexlab{}.
\newblock \showarticletitle{Lifting simplices to find injectivity}.
\newblock \bibinfo{journal}{\emph{ACM Transactions on Graphics (TOG)}}
  \bibinfo{volume}{39}, \bibinfo{number}{4} (\bibinfo{year}{2020}),
  \bibinfo{pages}{120--1}.
\newblock


\bibitem[Fulton et~al\mbox{.}(2019)]%
        {Fulton:LSD:2018}
\bibfield{author}{\bibinfo{person}{Lawson Fulton}, \bibinfo{person}{Vismay
  Modi}, \bibinfo{person}{David Duvenaud}, \bibinfo{person}{David I.~W. Levin},
  {and} \bibinfo{person}{Alec Jacobson}.} \bibinfo{year}{2019}\natexlab{}.
\newblock \showarticletitle{Latent-space Dynamics for Reduced Deformable
  Simulation}.
\newblock \bibinfo{journal}{\emph{Computer Graphics Forum}}
  (\bibinfo{year}{2019}).
\newblock


\bibitem[Gao et~al\mbox{.}(2018)]%
        {gaovcgan2018}
\bibfield{author}{\bibinfo{person}{Lin Gao}, \bibinfo{person}{Jie Yang},
  \bibinfo{person}{Yi-Ling Qiao}, \bibinfo{person}{Yu-Kun Lai},
  \bibinfo{person}{Paul~L Rosin}, \bibinfo{person}{Weiwei Xu}, {and}
  \bibinfo{person}{Shihong Xia}.} \bibinfo{year}{2018}\natexlab{}.
\newblock \showarticletitle{Automatic Unpaired Shape Deformation Transfer}.
\newblock \bibinfo{journal}{\emph{ACM Transactions on Graphics (Proceedings of
  ACM SIGGRAPH Asia 2018)}} \bibinfo{volume}{37}, \bibinfo{number}{6}
  (\bibinfo{year}{2018}), \bibinfo{pages}{To appear}.
\newblock


\bibitem[Gao et~al\mbox{.}(2019)]%
        {gao2019sdm}
\bibfield{author}{\bibinfo{person}{Lin Gao}, \bibinfo{person}{Jie Yang},
  \bibinfo{person}{Tong Wu}, \bibinfo{person}{Yu-Jie Yuan},
  \bibinfo{person}{Hongbo Fu}, \bibinfo{person}{Yu-Kun Lai}, {and}
  \bibinfo{person}{Hao Zhang}.} \bibinfo{year}{2019}\natexlab{}.
\newblock \showarticletitle{SDM-NET: Deep generative network for structured
  deformable mesh}.
\newblock \bibinfo{journal}{\emph{ACM Transactions on Graphics (TOG)}}
  \bibinfo{volume}{38}, \bibinfo{number}{6} (\bibinfo{year}{2019}),
  \bibinfo{pages}{1--15}.
\newblock


\bibitem[Groueix et~al\mbox{.}(2018a)]%
        {groueix2018coded}
\bibfield{author}{\bibinfo{person}{Thibault Groueix}, \bibinfo{person}{Matthew
  Fisher}, \bibinfo{person}{Vladimir~G. Kim}, \bibinfo{person}{Bryan~C.
  Russell}, {and} \bibinfo{person}{Mathieu Aubry}.}
  \bibinfo{year}{2018}\natexlab{a}.
\newblock \showarticletitle{{3D-CODED}: {3D} Correspondences by Deep
  Deformation}.
\newblock \bibinfo{journal}{\emph{ECCV}} (\bibinfo{year}{2018}).
\newblock


\bibitem[Groueix et~al\mbox{.}(2018b)]%
        {groueix2018atlasnet}
\bibfield{author}{\bibinfo{person}{Thibault Groueix}, \bibinfo{person}{Matthew
  Fisher}, \bibinfo{person}{Vladimir~G Kim}, \bibinfo{person}{Bryan~C Russell},
  {and} \bibinfo{person}{Mathieu Aubry}.} \bibinfo{year}{2018}\natexlab{b}.
\newblock \showarticletitle{AtlasNet: A Papier-Mache Approach to Learning 3D
  Surface Generation}.
\newblock \bibinfo{journal}{\emph{arXiv preprint arXiv:1802.05384}}
  (\bibinfo{year}{2018}).
\newblock


\bibitem[Groueix et~al\mbox{.}(2019)]%
        {Groueix19}
\bibfield{author}{\bibinfo{person}{Thibault Groueix}, \bibinfo{person}{Matthew
  Fisher}, \bibinfo{person}{Vladimir~G. Kim}, \bibinfo{person}{Bryan~C.
  Russell}, {and} \bibinfo{person}{Mathieu Aubry}.}
  \bibinfo{year}{2019}\natexlab{}.
\newblock \showarticletitle{Deep Self-Supervised Cycle-Consistent Deformation
  for Few-Shot Shape Segmentation}.
\newblock \bibinfo{journal}{\emph{SGP}} (\bibinfo{year}{2019}).
\newblock


\bibitem[Holden et~al\mbox{.}(2019)]%
        {holden2019subspace}
\bibfield{author}{\bibinfo{person}{Daniel Holden}, \bibinfo{person}{Bang~Chi
  Duong}, \bibinfo{person}{Sayantan Datta}, {and} \bibinfo{person}{Derek
  Nowrouzezahrai}.} \bibinfo{year}{2019}\natexlab{}.
\newblock \showarticletitle{Subspace neural physics: Fast data-driven
  interactive simulation}. In \bibinfo{booktitle}{\emph{Proceedings of the 18th
  annual ACM SIGGRAPH/Eurographics Symposium on Computer Animation}}.
  \bibinfo{pages}{1--12}.
\newblock


\bibitem[Holden et~al\mbox{.}(2015)]%
        {Holden:inverse_rig:2015}
\bibfield{author}{\bibinfo{person}{Daniel Holden}, \bibinfo{person}{Jun Saito},
  {and} \bibinfo{person}{Taku Komura}.} \bibinfo{year}{2015}\natexlab{}.
\newblock \showarticletitle{Learning an Inverse Rig Mapping for Character
  Animation}. In \bibinfo{booktitle}{\emph{Proceedings of the 14th ACM SIGGRAPH
  / Eurographics Symposium on Computer Animation}} (Los Angeles, California)
  \emph{(\bibinfo{series}{SCA '15})}. \bibinfo{publisher}{Association for
  Computing Machinery}, \bibinfo{address}{New York, NY, USA},
  \bibinfo{pages}{165–173}.
\newblock
\showISBNx{9781450334969}
\urldef\tempurl%
\url{https://doi.org/10.1145/2786784.2786788}
\showDOI{\tempurl}


\bibitem[Huang et~al\mbox{.}(2020)]%
        {huang2020meshode}
\bibfield{author}{\bibinfo{person}{Jingwei Huang}, \bibinfo{person}{Chiyu~Max
  Jiang}, \bibinfo{person}{Baiqiang Leng}, \bibinfo{person}{Bin Wang}, {and}
  \bibinfo{person}{Leonidas Guibas}.} \bibinfo{year}{2020}\natexlab{}.
\newblock \showarticletitle{Meshode: A robust and scalable framework for mesh
  deformation}.
\newblock \bibinfo{journal}{\emph{arXiv preprint arXiv:2005.11617}}
  (\bibinfo{year}{2020}).
\newblock


\bibitem[Jacobson et~al\mbox{.}(2011)]%
        {jacobson2011bounded}
\bibfield{author}{\bibinfo{person}{Alec Jacobson}, \bibinfo{person}{Ilya
  Baran}, \bibinfo{person}{Jovan Popovic}, {and} \bibinfo{person}{Olga
  Sorkine}.} \bibinfo{year}{2011}\natexlab{}.
\newblock \showarticletitle{Bounded biharmonic weights for real-time
  deformation.}
\newblock \bibinfo{journal}{\emph{ACM Trans. Graph.}} \bibinfo{volume}{30},
  \bibinfo{number}{4} (\bibinfo{year}{2011}), \bibinfo{pages}{78}.
\newblock


\bibitem[Jacobson et~al\mbox{.}(2014)]%
        {skinningcourse:2014}
\bibfield{author}{\bibinfo{person}{Alec Jacobson}, \bibinfo{person}{Zhigang
  Deng}, \bibinfo{person}{Ladislav Kavan}, {and} \bibinfo{person}{JP Lewis}.}
  \bibinfo{year}{2014}\natexlab{}.
\newblock \showarticletitle{Skinning: Real-time Shape Deformation}. In
  \bibinfo{booktitle}{\emph{ACM SIGGRAPH 2014 Courses}}.
\newblock


\bibitem[Jakab et~al\mbox{.}(2021)]%
        {jakab2021keypointdeformer}
\bibfield{author}{\bibinfo{person}{Tomas Jakab}, \bibinfo{person}{Richard
  Tucker}, \bibinfo{person}{Ameesh Makadia}, \bibinfo{person}{Jiajun Wu},
  \bibinfo{person}{Noah Snavely}, {and} \bibinfo{person}{Angjoo Kanazawa}.}
  \bibinfo{year}{2021}\natexlab{}.
\newblock \showarticletitle{KeypointDeformer: Unsupervised 3D Keypoint
  Discovery for Shape Control}. In \bibinfo{booktitle}{\emph{Proceedings of the
  IEEE/CVF Conference on Computer Vision and Pattern Recognition}}.
  \bibinfo{pages}{12783--12792}.
\newblock


\bibitem[Jiang et~al\mbox{.}(2020)]%
        {jiang2020shapeflow}
\bibfield{author}{\bibinfo{person}{Chiyu Jiang}, \bibinfo{person}{Jingwei
  Huang}, \bibinfo{person}{Andrea Tagliasacchi}, \bibinfo{person}{Leonidas
  Guibas}, {et~al\mbox{.}}} \bibinfo{year}{2020}\natexlab{}.
\newblock \showarticletitle{Shapeflow: Learnable deformations among 3d shapes}.
\newblock \bibinfo{journal}{\emph{arXiv preprint arXiv:2006.07982}}
  (\bibinfo{year}{2020}).
\newblock


\bibitem[Ju et~al\mbox{.}(2005)]%
        {ju2005mean}
\bibfield{author}{\bibinfo{person}{Tao Ju}, \bibinfo{person}{Scott Schaefer},
  {and} \bibinfo{person}{Joe Warren}.} \bibinfo{year}{2005}\natexlab{}.
\newblock \showarticletitle{Mean value coordinates for closed triangular
  meshes}.
\newblock In \bibinfo{booktitle}{\emph{ACM Siggraph 2005 Papers}}.
  \bibinfo{pages}{561--566}.
\newblock


\bibitem[Kanazawa et~al\mbox{.}(2016)]%
        {kanazawa2016learning}
\bibfield{author}{\bibinfo{person}{Angjoo Kanazawa}, \bibinfo{person}{Shahar
  Kovalsky}, \bibinfo{person}{Ronen Basri}, {and} \bibinfo{person}{David
  Jacobs}.} \bibinfo{year}{2016}\natexlab{}.
\newblock \showarticletitle{Learning 3d deformation of animals from 2d images}.
  In \bibinfo{booktitle}{\emph{Computer Graphics Forum}},
  Vol.~\bibinfo{volume}{35}. Wiley Online Library, \bibinfo{pages}{365--374}.
\newblock


\bibitem[Kanazawa et~al\mbox{.}(2018)]%
        {cmrKanazawa18}
\bibfield{author}{\bibinfo{person}{Angjoo Kanazawa}, \bibinfo{person}{Shubham
  Tulsiani}, \bibinfo{person}{Alexei~A. Efros}, {and} \bibinfo{person}{Jitendra
  Malik}.} \bibinfo{year}{2018}\natexlab{}.
\newblock \showarticletitle{Learning Category-Specific Mesh Reconstruction from
  Image Collections}. In \bibinfo{booktitle}{\emph{ECCV}}.
\newblock


\bibitem[Kavan et~al\mbox{.}(2008)]%
        {kavan2008geometric}
\bibfield{author}{\bibinfo{person}{Ladislav Kavan}, \bibinfo{person}{Steven
  Collins}, \bibinfo{person}{Ji{\v{r}}{\'\i} {\v{Z}}{\'a}ra}, {and}
  \bibinfo{person}{Carol O'Sullivan}.} \bibinfo{year}{2008}\natexlab{}.
\newblock \showarticletitle{Geometric skinning with approximate dual quaternion
  blending}.
\newblock \bibinfo{journal}{\emph{ACM Transactions on Graphics (TOG)}}
  \bibinfo{volume}{27}, \bibinfo{number}{4} (\bibinfo{year}{2008}),
  \bibinfo{pages}{1--23}.
\newblock


\bibitem[Kim and Eberle(2020)]%
        {Kim:DynamicDeformables:2020}
\bibfield{author}{\bibinfo{person}{Theodore Kim} {and} \bibinfo{person}{David
  Eberle}.} \bibinfo{year}{2020}\natexlab{}.
\newblock \showarticletitle{Dynamic Deformables: Implementation and Production
  Practicalities}. In \bibinfo{booktitle}{\emph{ACM SIGGRAPH 2020 Courses}}
  (Virtual Event, USA) \emph{(\bibinfo{series}{SIGGRAPH '20})}.
  \bibinfo{publisher}{Association for Computing Machinery},
  \bibinfo{address}{New York, NY, USA}, Article \bibinfo{articleno}{23},
  \bibinfo{numpages}{182}~pages.
\newblock
\showISBNx{9781450379724}
\urldef\tempurl%
\url{https://doi.org/10.1145/3388769.3407490}
\showDOI{\tempurl}


\bibitem[Kingma and Ba(2015)]%
        {KingmaAdam}
\bibfield{author}{\bibinfo{person}{Diederik~P. Kingma} {and}
  \bibinfo{person}{Jimmy Ba}.} \bibinfo{year}{2015}\natexlab{}.
\newblock \showarticletitle{Adam: {A} Method for Stochastic Optimization}. In
  \bibinfo{booktitle}{\emph{3rd International Conference on Learning
  Representations, {ICLR} 2015, San Diego, CA, USA, May 7-9, 2015, Conference
  Track Proceedings}}, \bibfield{editor}{\bibinfo{person}{Yoshua Bengio} {and}
  \bibinfo{person}{Yann LeCun}} (Eds.).
\newblock
\urldef\tempurl%
\url{http://arxiv.org/abs/1412.6980}
\showURL{%
\tempurl}


\bibitem[Kovalsky et~al\mbox{.}(2014)]%
        {kovalsky2014controlling}
\bibfield{author}{\bibinfo{person}{Shahar~Z Kovalsky}, \bibinfo{person}{Noam
  Aigerman}, \bibinfo{person}{Ronen Basri}, {and} \bibinfo{person}{Yaron
  Lipman}.} \bibinfo{year}{2014}\natexlab{}.
\newblock \showarticletitle{Controlling singular values with semidefinite
  programming.}
\newblock \bibinfo{journal}{\emph{ACM Trans. Graph.}} \bibinfo{volume}{33},
  \bibinfo{number}{4} (\bibinfo{year}{2014}), \bibinfo{pages}{68--1}.
\newblock


\bibitem[L\'evy et~al\mbox{.}(2002)]%
        {levy2002lscm}
\bibfield{author}{\bibinfo{person}{Bruno L\'evy}, \bibinfo{person}{Sylvain
  Petitjean}, \bibinfo{person}{Nicolas Ray}, {and}
  \bibinfo{person}{J\'er\^{o}me Maillot}.} \bibinfo{year}{2002}\natexlab{}.
\newblock \showarticletitle{Least Squares Conformal Maps for Automatic Texture
  Atlas Generation}. In \bibinfo{booktitle}{\emph{SIGGRAPH}}.
\newblock


\bibitem[Li et~al\mbox{.}(2021)]%
        {li2021learning}
\bibfield{author}{\bibinfo{person}{Peizhuo Li}, \bibinfo{person}{Kfir Aberman},
  \bibinfo{person}{Rana Hanocka}, \bibinfo{person}{Libin Liu},
  \bibinfo{person}{Olga Sorkine-Hornung}, {and} \bibinfo{person}{Baoquan
  Chen}.} \bibinfo{year}{2021}\natexlab{}.
\newblock \showarticletitle{Learning Skeletal Articulations with Neural Blend
  Shapes}.
\newblock \bibinfo{journal}{\emph{ACM Transactions on Graphics (TOG)}}
  \bibinfo{volume}{40}, \bibinfo{number}{4} (\bibinfo{year}{2021}),
  \bibinfo{pages}{1}.
\newblock


\bibitem[Lipman(2012)]%
        {lipman2012bounded}
\bibfield{author}{\bibinfo{person}{Yaron Lipman}.}
  \bibinfo{year}{2012}\natexlab{}.
\newblock \showarticletitle{Bounded distortion mapping spaces for triangular
  meshes}.
\newblock \bibinfo{journal}{\emph{ACM Transactions on Graphics (TOG)}}
  \bibinfo{volume}{31}, \bibinfo{number}{4} (\bibinfo{year}{2012}),
  \bibinfo{pages}{1--13}.
\newblock


\bibitem[Lipman et~al\mbox{.}(2008)]%
        {lipman2008green}
\bibfield{author}{\bibinfo{person}{Yaron Lipman}, \bibinfo{person}{David
  Levin}, {and} \bibinfo{person}{Daniel Cohen-Or}.}
  \bibinfo{year}{2008}\natexlab{}.
\newblock \showarticletitle{Green coordinates}.
\newblock \bibinfo{journal}{\emph{ACM Transactions on Graphics (TOG)}}
  \bibinfo{volume}{27}, \bibinfo{number}{3} (\bibinfo{year}{2008}),
  \bibinfo{pages}{1--10}.
\newblock


\bibitem[Lipman et~al\mbox{.}(2004)]%
        {lipman2004differential}
\bibfield{author}{\bibinfo{person}{Yaron Lipman}, \bibinfo{person}{Olga
  Sorkine}, \bibinfo{person}{Daniel Cohen-Or}, \bibinfo{person}{David Levin},
  \bibinfo{person}{Christian Rossi}, {and} \bibinfo{person}{Hans-Peter
  Seidel}.} \bibinfo{year}{2004}\natexlab{}.
\newblock \showarticletitle{Differential coordinates for interactive mesh
  editing}. In \bibinfo{booktitle}{\emph{Proceedings Shape Modeling
  Applications, 2004.}} IEEE, \bibinfo{pages}{181--190}.
\newblock


\bibitem[Litany et~al\mbox{.}(2018)]%
        {litany2018deformable}
\bibfield{author}{\bibinfo{person}{Or Litany}, \bibinfo{person}{Alex
  Bronstein}, \bibinfo{person}{Michael Bronstein}, {and}
  \bibinfo{person}{Ameesh Makadia}.} \bibinfo{year}{2018}\natexlab{}.
\newblock \showarticletitle{Deformable shape completion with graph
  convolutional autoencoders}. In \bibinfo{booktitle}{\emph{Proceedings of the
  IEEE conference on computer vision and pattern recognition}}.
  \bibinfo{pages}{1886--1895}.
\newblock


\bibitem[Liu et~al\mbox{.}(2008a)]%
        {Ligang:arap:2008}
\bibfield{author}{\bibinfo{person}{Ligang Liu}, \bibinfo{person}{Lei Zhang},
  \bibinfo{person}{Yin Xu}, \bibinfo{person}{Craig Gotsman}, {and}
  \bibinfo{person}{Steven~J. Gortler}.} \bibinfo{year}{2008}\natexlab{a}.
\newblock \showarticletitle{A Local/Global Approach to Mesh Parameterization}.
  In \bibinfo{booktitle}{\emph{Proceedings of the Symposium on Geometry
  Processing}} (Copenhagen, Denmark) \emph{(\bibinfo{series}{SGP '08})}.
  \bibinfo{publisher}{Eurographics Association}, \bibinfo{address}{Goslar,
  DEU}, \bibinfo{pages}{1495–1504}.
\newblock


\bibitem[Liu et~al\mbox{.}(2008b)]%
        {liu2008local}
\bibfield{author}{\bibinfo{person}{Ligang Liu}, \bibinfo{person}{Lei Zhang},
  \bibinfo{person}{Yin Xu}, \bibinfo{person}{Craig Gotsman}, {and}
  \bibinfo{person}{Steven~J Gortler}.} \bibinfo{year}{2008}\natexlab{b}.
\newblock \showarticletitle{A local/global approach to mesh parameterization}.
  In \bibinfo{booktitle}{\emph{Computer Graphics Forum}},
  Vol.~\bibinfo{volume}{27}. Wiley Online Library, \bibinfo{pages}{1495--1504}.
\newblock


\bibitem[Liu et~al\mbox{.}(2019)]%
        {liu2019neuroskinning}
\bibfield{author}{\bibinfo{person}{Lijuan Liu}, \bibinfo{person}{Youyi Zheng},
  \bibinfo{person}{Di Tang}, \bibinfo{person}{Yi Yuan},
  \bibinfo{person}{Changjie Fan}, {and} \bibinfo{person}{Kun Zhou}.}
  \bibinfo{year}{2019}\natexlab{}.
\newblock \showarticletitle{NeuroSkinning: Automatic skin binding for
  production characters with deep graph networks}.
\newblock \bibinfo{journal}{\emph{ACM Transactions on Graphics (TOG)}}
  \bibinfo{volume}{38}, \bibinfo{number}{4} (\bibinfo{year}{2019}),
  \bibinfo{pages}{1--12}.
\newblock


\bibitem[Liu et~al\mbox{.}(2021)]%
        {liu2021deepmetahandles}
\bibfield{author}{\bibinfo{person}{Minghua Liu}, \bibinfo{person}{Minhyuk
  Sung}, \bibinfo{person}{Radomir Mech}, {and} \bibinfo{person}{Hao Su}.}
  \bibinfo{year}{2021}\natexlab{}.
\newblock \showarticletitle{DeepMetaHandles: Learning Deformation Meta-Handles
  of 3D Meshes with Biharmonic Coordinates}. In
  \bibinfo{booktitle}{\emph{Proceedings of the IEEE/CVF Conference on Computer
  Vision and Pattern Recognition}}. \bibinfo{pages}{12--21}.
\newblock


\bibitem[Morreale et~al\mbox{.}(2021)]%
        {morreale2021neural}
\bibfield{author}{\bibinfo{person}{Luca Morreale}, \bibinfo{person}{Noam
  Aigerman}, \bibinfo{person}{Vladimir~G Kim}, {and} \bibinfo{person}{Niloy~J
  Mitra}.} \bibinfo{year}{2021}\natexlab{}.
\newblock \showarticletitle{Neural Surface Maps}. In
  \bibinfo{booktitle}{\emph{Proceedings of the IEEE/CVF Conference on Computer
  Vision and Pattern Recognition}}. \bibinfo{pages}{4639--4648}.
\newblock


\bibitem[Myles and Zorin(2013)]%
        {myles2013controlled}
\bibfield{author}{\bibinfo{person}{Ashish Myles} {and} \bibinfo{person}{Denis
  Zorin}.} \bibinfo{year}{2013}\natexlab{}.
\newblock \showarticletitle{Controlled-distortion constrained global
  parametrization}.
\newblock \bibinfo{journal}{\emph{ACM Transactions on Graphics (TOG)}}
  \bibinfo{volume}{32}, \bibinfo{number}{4} (\bibinfo{year}{2013}),
  \bibinfo{pages}{1--14}.
\newblock


\bibitem[Okuta et~al\mbox{.}(2017)]%
        {cupy_learningsys2017}
\bibfield{author}{\bibinfo{person}{Ryosuke Okuta}, \bibinfo{person}{Yuya Unno},
  \bibinfo{person}{Daisuke Nishino}, \bibinfo{person}{Shohei Hido}, {and}
  \bibinfo{person}{Crissman Loomis}.} \bibinfo{year}{2017}\natexlab{}.
\newblock \showarticletitle{CuPy: A NumPy-Compatible Library for NVIDIA GPU
  Calculations}. In \bibinfo{booktitle}{\emph{Proceedings of Workshop on
  Machine Learning Systems (LearningSys) in The Thirty-first Annual Conference
  on Neural Information Processing Systems (NIPS)}}.
\newblock
\urldef\tempurl%
\url{http://learningsys.org/nips17/assets/papers/paper_16.pdf}
\showURL{%
\tempurl}


\bibitem[Osman et~al\mbox{.}(2020)]%
        {STAR:2020}
\bibfield{author}{\bibinfo{person}{Ahmed A~A Osman}, \bibinfo{person}{Timo
  Bolkart}, {and} \bibinfo{person}{Michael~J. Black}.}
  \bibinfo{year}{2020}\natexlab{}.
\newblock \showarticletitle{{STAR}: A Sparse Trained Articulated Human Body
  Regressor}. In \bibinfo{booktitle}{\emph{European Conference on Computer
  Vision (ECCV)}}. \bibinfo{pages}{598--613}.
\newblock
\urldef\tempurl%
\url{https://star.is.tue.mpg.de}
\showURL{%
\tempurl}


\bibitem[Park et~al\mbox{.}(2019)]%
        {park2019deepsdf}
\bibfield{author}{\bibinfo{person}{Jeong~Joon Park}, \bibinfo{person}{Peter
  Florence}, \bibinfo{person}{Julian Straub}, \bibinfo{person}{Richard~A.
  Newcombe}, {and} \bibinfo{person}{Steven Lovegrove}.}
  \bibinfo{year}{2019}\natexlab{}.
\newblock \showarticletitle{DeepSDF: Learning Continuous Signed Distance
  Functions for Shape Representation}.
\newblock \bibinfo{journal}{\emph{CVPR}} (\bibinfo{year}{2019}).
\newblock


\bibitem[Paszke et~al\mbox{.}(2019)]%
        {NEURIPS2019_9015}
\bibfield{author}{\bibinfo{person}{Adam Paszke}, \bibinfo{person}{Sam Gross},
  \bibinfo{person}{Francisco Massa}, \bibinfo{person}{Adam Lerer},
  \bibinfo{person}{James Bradbury}, \bibinfo{person}{Gregory Chanan},
  \bibinfo{person}{Trevor Killeen}, \bibinfo{person}{Zeming Lin},
  \bibinfo{person}{Natalia Gimelshein}, \bibinfo{person}{Luca Antiga},
  \bibinfo{person}{Alban Desmaison}, \bibinfo{person}{Andreas Kopf},
  \bibinfo{person}{Edward Yang}, \bibinfo{person}{Zachary DeVito},
  \bibinfo{person}{Martin Raison}, \bibinfo{person}{Alykhan Tejani},
  \bibinfo{person}{Sasank Chilamkurthy}, \bibinfo{person}{Benoit Steiner},
  \bibinfo{person}{Lu Fang}, \bibinfo{person}{Junjie Bai}, {and}
  \bibinfo{person}{Soumith Chintala}.} \bibinfo{year}{2019}\natexlab{}.
\newblock \showarticletitle{PyTorch: An Imperative Style, High-Performance Deep
  Learning Library}.
\newblock In \bibinfo{booktitle}{\emph{Advances in Neural Information
  Processing Systems 32}}, \bibfield{editor}{\bibinfo{person}{H.~Wallach},
  \bibinfo{person}{H.~Larochelle}, \bibinfo{person}{A.~Beygelzimer},
  \bibinfo{person}{F.~d\textquotesingle Alch\'{e}-Buc},
  \bibinfo{person}{E.~Fox}, {and} \bibinfo{person}{R.~Garnett}} (Eds.).
  \bibinfo{publisher}{Curran Associates, Inc.}, \bibinfo{pages}{8024--8035}.
\newblock
\urldef\tempurl%
\url{http://papers.neurips.cc/paper/9015-pytorch-an-imperative-style-high-performance-deep-learning-library.pdf}
\showURL{%
\tempurl}


\bibitem[Pinkall and Polthier(1993)]%
        {Pinkall93computingdiscrete}
\bibfield{author}{\bibinfo{person}{Ulrich Pinkall} {and}
  \bibinfo{person}{Konrad Polthier}.} \bibinfo{year}{1993}\natexlab{}.
\newblock \showarticletitle{Computing Discrete Minimal Surfaces and Their
  Conjugates}.
\newblock \bibinfo{journal}{\emph{EXPERIMENTAL MATHEMATICS}}
  \bibinfo{volume}{2} (\bibinfo{year}{1993}), \bibinfo{pages}{15--36}.
\newblock


\bibitem[Qi et~al\mbox{.}(2017)]%
        {qi2017pointnet}
\bibfield{author}{\bibinfo{person}{Charles~R Qi}, \bibinfo{person}{Hao Su},
  \bibinfo{person}{Kaichun Mo}, {and} \bibinfo{person}{Leonidas~J Guibas}.}
  \bibinfo{year}{2017}\natexlab{}.
\newblock \showarticletitle{Pointnet: Deep learning on point sets for 3d
  classification and segmentation}. In \bibinfo{booktitle}{\emph{Proceedings of
  the IEEE conference on computer vision and pattern recognition}}.
  \bibinfo{pages}{652--660}.
\newblock


\bibitem[Rabinovich et~al\mbox{.}(2017)]%
        {Rabinovich:SLIM:2017}
\bibfield{author}{\bibinfo{person}{Michael Rabinovich}, \bibinfo{person}{Roi
  Poranne}, \bibinfo{person}{Daniele Panozzo}, {and} \bibinfo{person}{Olga
  Sorkine-Hornung}.} \bibinfo{year}{2017}\natexlab{}.
\newblock \showarticletitle{Scalable Locally Injective Mappings}.
\newblock \bibinfo{journal}{\emph{ACM Transactions on Graphics}}
  \bibinfo{volume}{36}, \bibinfo{number}{2} (\bibinfo{date}{April}
  \bibinfo{year}{2017}), \bibinfo{pages}{16:1--16:16}.
\newblock


\bibitem[Romero et~al\mbox{.}(2021)]%
        {Romero:2021}
\bibfield{author}{\bibinfo{person}{Cristian Romero}, \bibinfo{person}{Dan
  Casas}, \bibinfo{person}{Jesus Perez}, {and} \bibinfo{person}{Miguel~A.
  Otaduy}.} \bibinfo{year}{2021}\natexlab{}.
\newblock \showarticletitle{Learning Contact Corrections for Handle-Based
  Subspace Dynamics}.
\newblock \bibinfo{journal}{\emph{ACM Trans. on Graphics (Proc. of ACM
  SIGGRAPH)}} \bibinfo{volume}{40}, \bibinfo{number}{4} (\bibinfo{year}{2021}).
\newblock
\urldef\tempurl%
\url{http://gmrv.es/Publications/2021/RCPO21}
\showURL{%
\tempurl}


\bibitem[Sahillio{\u{g}}lu(2020)]%
        {sahilliouglu2020recent}
\bibfield{author}{\bibinfo{person}{Yusuf Sahillio{\u{g}}lu}.}
  \bibinfo{year}{2020}\natexlab{}.
\newblock \showarticletitle{Recent advances in shape correspondence}.
\newblock \bibinfo{journal}{\emph{The Visual Computer}} \bibinfo{volume}{36},
  \bibinfo{number}{8} (\bibinfo{year}{2020}), \bibinfo{pages}{1705--1721}.
\newblock


\bibitem[Sch{\"u}ller et~al\mbox{.}(2013)]%
        {schuller2013locally}
\bibfield{author}{\bibinfo{person}{Christian Sch{\"u}ller},
  \bibinfo{person}{Ladislav Kavan}, \bibinfo{person}{Daniele Panozzo}, {and}
  \bibinfo{person}{Olga Sorkine-Hornung}.} \bibinfo{year}{2013}\natexlab{}.
\newblock \showarticletitle{Locally injective mappings}. In
  \bibinfo{booktitle}{\emph{Computer Graphics Forum}},
  Vol.~\bibinfo{volume}{32}. Wiley Online Library, \bibinfo{pages}{125--135}.
\newblock


\bibitem[Sharp et~al\mbox{.}(2022)]%
        {sharp2022diffusionnet}
\bibfield{author}{\bibinfo{person}{Nicholas Sharp}, \bibinfo{person}{Souhaib
  Attaiki}, \bibinfo{person}{Keenan Crane}, {and} \bibinfo{person}{Maks
  Ovsjanikov}.} \bibinfo{year}{2022}\natexlab{}.
\newblock \showarticletitle{Diffusionnet: Discretization agnostic learning on
  surfaces}.
\newblock \bibinfo{journal}{\emph{ACM Transactions on Graphics (TOG)}}
  \bibinfo{volume}{41}, \bibinfo{number}{3} (\bibinfo{year}{2022}),
  \bibinfo{pages}{1--16}.
\newblock


\bibitem[Sheffer et~al\mbox{.}(2007)]%
        {Sheffer:meshparam:2007}
\bibfield{author}{\bibinfo{person}{Alla Sheffer}, \bibinfo{person}{K Hormann},
  \bibinfo{person}{B Levy}, \bibinfo{person}{M Desbrun}, \bibinfo{person}{K
  Zhou}, \bibinfo{person}{E Praun}, {and} \bibinfo{person}{H Hoppe}.}
  \bibinfo{year}{2007}\natexlab{}.
\newblock \showarticletitle{Mesh parameterization: Theory and practice}.
\newblock \bibinfo{journal}{\emph{ACM SIGGRAPPH, course notes}}
  \bibinfo{volume}{10}, \bibinfo{number}{1281500.1281510}
  (\bibinfo{year}{2007}).
\newblock


\bibitem[Shen et~al\mbox{.}(2021)]%
        {Shen:2021}
\bibfield{author}{\bibinfo{person}{Siyuan Shen}, \bibinfo{person}{Yin Yang},
  \bibinfo{person}{Tianjia Shao}, \bibinfo{person}{He Wang},
  \bibinfo{person}{Chenfanfu Jiang}, \bibinfo{person}{Lei Lan}, {and}
  \bibinfo{person}{Kun Zhou}.} \bibinfo{year}{2021}\natexlab{}.
\newblock \showarticletitle{High-order differentiable autoencoder for nonlinear
  model reduction}.
\newblock \bibinfo{journal}{\emph{ACM Transactions on Graphics}}.
\newblock


\bibitem[Smith and Schaefer(2015)]%
        {Smith:bijective:2015}
\bibfield{author}{\bibinfo{person}{Jason Smith} {and} \bibinfo{person}{Scott
  Schaefer}.} \bibinfo{year}{2015}\natexlab{}.
\newblock \showarticletitle{Bijective Parameterization with Free Boundaries}.
\newblock \bibinfo{journal}{\emph{ACM Trans. Graph.}} \bibinfo{volume}{34},
  \bibinfo{number}{4}, Article \bibinfo{articleno}{70} (\bibinfo{date}{jul}
  \bibinfo{year}{2015}), \bibinfo{numpages}{9}~pages.
\newblock
\showISSN{0730-0301}
\urldef\tempurl%
\url{https://doi.org/10.1145/2766947}
\showDOI{\tempurl}


\bibitem[Sorkine and Alexa(2007)]%
        {sorkine2007arap}
\bibfield{author}{\bibinfo{person}{Olga Sorkine} {and} \bibinfo{person}{Marc
  Alexa}.} \bibinfo{year}{2007}\natexlab{}.
\newblock \showarticletitle{As-Rigid-As-Possible Surface Modeling}. In
  \bibinfo{booktitle}{\emph{SGP}}.
\newblock


\bibitem[Sorkine and Botsch(2009)]%
        {DeformationTutorial:2009}
\bibfield{author}{\bibinfo{person}{Olga Sorkine} {and} \bibinfo{person}{Mario
  Botsch}.} \bibinfo{year}{2009}\natexlab{}.
\newblock \showarticletitle{Interactive Shape Modeling and Deformation}. In
  \bibinfo{booktitle}{\emph{EUROGRAPHICS Tutorials}}.
\newblock


\bibitem[Sorkine et~al\mbox{.}(2004)]%
        {sorkine2004laplacian}
\bibfield{author}{\bibinfo{person}{Olga Sorkine}, \bibinfo{person}{Daniel
  Cohen-Or}, \bibinfo{person}{Yaron Lipman}, \bibinfo{person}{Marc Alexa},
  \bibinfo{person}{Christian R{\"o}ssl}, {and} \bibinfo{person}{H-P Seidel}.}
  \bibinfo{year}{2004}\natexlab{}.
\newblock \showarticletitle{Laplacian surface editing}. In
  \bibinfo{booktitle}{\emph{Proceedings of the 2004 Eurographics/ACM SIGGRAPH
  symposium on Geometry processing}}. \bibinfo{pages}{175--184}.
\newblock


\bibitem[Sumner and Popovi{\'c}(2004)]%
        {sumner2004deformation}
\bibfield{author}{\bibinfo{person}{Robert~W Sumner} {and}
  \bibinfo{person}{Jovan Popovi{\'c}}.} \bibinfo{year}{2004}\natexlab{}.
\newblock \showarticletitle{Deformation transfer for triangle meshes}.
\newblock \bibinfo{journal}{\emph{ACM Transactions on graphics (TOG)}}
  \bibinfo{volume}{23}, \bibinfo{number}{3} (\bibinfo{year}{2004}),
  \bibinfo{pages}{399--405}.
\newblock


\bibitem[Sun et~al\mbox{.}(2021)]%
        {Sun:2021}
\bibfield{author}{\bibinfo{person}{Bo Sun}, \bibinfo{person}{Xiangru Huang},
  \bibinfo{person}{Qixing Huang}, \bibinfo{person}{Zaiwei Zhang},
  \bibinfo{person}{Junfeng Jiang}, {and} \bibinfo{person}{Chandrajit Bajaj}.}
  \bibinfo{year}{2021}\natexlab{}.
\newblock \showarticletitle{ARAPReg: An As-Rigid-As Possible Regularization
  Loss for Learning Deformable Shape Generators}. In
  \bibinfo{booktitle}{\emph{ICCV}}.
\newblock


\bibitem[Tan et~al\mbox{.}(2018)]%
        {tan2018meshvae}
\bibfield{author}{\bibinfo{person}{Qingyang Tan}, \bibinfo{person}{Lin Gao},
  \bibinfo{person}{Yu-Kun Lai}, {and} \bibinfo{person}{Shihong Xia}.}
  \bibinfo{year}{2018}\natexlab{}.
\newblock \showarticletitle{Variational Autoencoders for Deforming {3D} Mesh
  Models}. In \bibinfo{booktitle}{\emph{CVPR}}.
\newblock


\bibitem[Tarini et~al\mbox{.}(2004)]%
        {Tarini:polycube:2004}
\bibfield{author}{\bibinfo{person}{Marco Tarini}, \bibinfo{person}{Kai
  Hormann}, \bibinfo{person}{Paolo Cignoni}, {and} \bibinfo{person}{Claudio
  Montani}.} \bibinfo{year}{2004}\natexlab{}.
\newblock \showarticletitle{PolyCube-Maps}. In \bibinfo{booktitle}{\emph{ACM
  SIGGRAPH 2004 Papers}} (Los Angeles, California)
  \emph{(\bibinfo{series}{SIGGRAPH '04})}. \bibinfo{publisher}{Association for
  Computing Machinery}, \bibinfo{address}{New York, NY, USA},
  \bibinfo{pages}{853–860}.
\newblock
\showISBNx{9781450378239}
\urldef\tempurl%
\url{https://doi.org/10.1145/1186562.1015810}
\showDOI{\tempurl}


\bibitem[Uy et~al\mbox{.}(2020)]%
        {uy2020deformation}
\bibfield{author}{\bibinfo{person}{Mikaela~Angelina Uy},
  \bibinfo{person}{Jingwei Huang}, \bibinfo{person}{Minhyuk Sung},
  \bibinfo{person}{Tolga Birdal}, {and} \bibinfo{person}{Leonidas Guibas}.}
  \bibinfo{year}{2020}\natexlab{}.
\newblock \showarticletitle{Deformation-aware 3d model embedding and
  retrieval}. In \bibinfo{booktitle}{\emph{European Conference on Computer
  Vision}}. Springer, \bibinfo{pages}{397--413}.
\newblock


\bibitem[Varol et~al\mbox{.}(2017)]%
        {varol17_surreal}
\bibfield{author}{\bibinfo{person}{G{\"u}l Varol}, \bibinfo{person}{Javier
  Romero}, \bibinfo{person}{Xavier Martin}, \bibinfo{person}{Naureen Mahmood},
  \bibinfo{person}{Michael~J. Black}, \bibinfo{person}{Ivan Laptev}, {and}
  \bibinfo{person}{Cordelia Schmid}.} \bibinfo{year}{2017}\natexlab{}.
\newblock \showarticletitle{Learning from Synthetic Humans}. In
  \bibinfo{booktitle}{\emph{CVPR}}.
\newblock


\bibitem[Wang et~al\mbox{.}(2018)]%
        {wang2018pixel2mesh}
\bibfield{author}{\bibinfo{person}{Nanyang Wang}, \bibinfo{person}{Yinda
  Zhang}, \bibinfo{person}{Zhuwen Li}, \bibinfo{person}{Yanwei Fu},
  \bibinfo{person}{Wei Liu}, {and} \bibinfo{person}{Yu-Gang Jiang}.}
  \bibinfo{year}{2018}\natexlab{}.
\newblock \showarticletitle{Pixel2mesh: Generating 3d mesh models from single
  rgb images}. In \bibinfo{booktitle}{\emph{Proceedings of the European
  Conference on Computer Vision (ECCV)}}. \bibinfo{pages}{52--67}.
\newblock


\bibitem[Weber and Zorin(2014)]%
        {weber2014locally}
\bibfield{author}{\bibinfo{person}{Ofir Weber} {and} \bibinfo{person}{Denis
  Zorin}.} \bibinfo{year}{2014}\natexlab{}.
\newblock \showarticletitle{Locally injective parametrization with arbitrary
  fixed boundaries}.
\newblock \bibinfo{journal}{\emph{ACM Transactions on Graphics (TOG)}}
  \bibinfo{volume}{33}, \bibinfo{number}{4} (\bibinfo{year}{2014}),
  \bibinfo{pages}{1--12}.
\newblock


\bibitem[Williams et~al\mbox{.}(2019)]%
        {williams2019deep}
\bibfield{author}{\bibinfo{person}{Francis Williams}, \bibinfo{person}{Teseo
  Schneider}, \bibinfo{person}{Claudio Silva}, \bibinfo{person}{Denis Zorin},
  \bibinfo{person}{Joan Bruna}, {and} \bibinfo{person}{Daniele Panozzo}.}
  \bibinfo{year}{2019}\natexlab{}.
\newblock \showarticletitle{Deep geometric prior for surface reconstruction}.
  In \bibinfo{booktitle}{\emph{Proceedings of the IEEE/CVF Conference on
  Computer Vision and Pattern Recognition}}. \bibinfo{pages}{10130--10139}.
\newblock


\bibitem[Wu and He(2018)]%
        {wu2018group}
\bibfield{author}{\bibinfo{person}{Yuxin Wu} {and} \bibinfo{person}{Kaiming
  He}.} \bibinfo{year}{2018}\natexlab{}.
\newblock \showarticletitle{Group normalization}. In
  \bibinfo{booktitle}{\emph{Proceedings of the European conference on computer
  vision (ECCV)}}. \bibinfo{pages}{3--19}.
\newblock


\bibitem[Xu et~al\mbox{.}(2020)]%
        {RigNet}
\bibfield{author}{\bibinfo{person}{Zhan Xu}, \bibinfo{person}{Yang Zhou},
  \bibinfo{person}{Evangelos Kalogerakis}, \bibinfo{person}{Chris Landreth},
  {and} \bibinfo{person}{Karan Singh}.} \bibinfo{year}{2020}\natexlab{}.
\newblock \showarticletitle{RigNet: Neural Rigging for Articulated Characters}.
\newblock \bibinfo{journal}{\emph{ACM Trans. on Graphics}}
  \bibinfo{volume}{39} (\bibinfo{year}{2020}).
\newblock


\bibitem[Xu et~al\mbox{.}(2019)]%
        {AnimSkelVolNet}
\bibfield{author}{\bibinfo{person}{Zhan Xu}, \bibinfo{person}{Yang Zhou},
  \bibinfo{person}{Evangelos Kalogerakis}, {and} \bibinfo{person}{Karan
  Singh}.} \bibinfo{year}{2019}\natexlab{}.
\newblock \showarticletitle{Predicting Animation Skeletons for 3D Articulated
  Models via Volumetric Nets}. In \bibinfo{booktitle}{\emph{2019 International
  Conference on 3D Vision (3DV)}}.
\newblock


\bibitem[Yang et~al\mbox{.}(2021)]%
        {Yang:2021}
\bibfield{author}{\bibinfo{person}{Guandao Yang}, \bibinfo{person}{Serge
  Belongie}, \bibinfo{person}{Bharath Hariharan}, {and}
  \bibinfo{person}{Vladlen Koltun}.} \bibinfo{year}{2021}\natexlab{}.
\newblock \showarticletitle{Geometry Processing with Neural Fields}.
\newblock \bibinfo{journal}{\emph{NeurIPS}}.
\newblock


\bibitem[Yifan et~al\mbox{.}(2020)]%
        {wang2019neural}
\bibfield{author}{\bibinfo{person}{Wang Yifan}, \bibinfo{person}{Noam
  Aigerman}, \bibinfo{person}{Vladimir~G. Kim}, \bibinfo{person}{Siddhartha
  Chaudhuri}, {and} \bibinfo{person}{Olga Sorkine-Hornung}.}
  \bibinfo{year}{2020}\natexlab{}.
\newblock \showarticletitle{Neural Cages for Detail-Preserving {3D}
  Deformations}. In \bibinfo{booktitle}{\emph{CVPR}}.
\newblock


\bibitem[Yin et~al\mbox{.}(2021)]%
        {yin2021_3DStyleNet}
\bibfield{author}{\bibinfo{person}{Kangxue Yin}, \bibinfo{person}{Jun Gao},
  \bibinfo{person}{Maria Shugrina}, \bibinfo{person}{Sameh Khamis}, {and}
  \bibinfo{person}{Sanja Fidler}.} \bibinfo{year}{2021}\natexlab{}.
\newblock \showarticletitle{3DStyleNet: Creating 3D Shapes with Geometric and
  Texture Style Variations}. In \bibinfo{booktitle}{\emph{Proceedings of
  International Conference on Computer Vision (ICCV)}}.
\newblock


\bibitem[Yu et~al\mbox{.}(2004)]%
        {yu2004mesh}
\bibfield{author}{\bibinfo{person}{Yizhou Yu}, \bibinfo{person}{Kun Zhou},
  \bibinfo{person}{Dong Xu}, \bibinfo{person}{Xiaohan Shi},
  \bibinfo{person}{Hujun Bao}, \bibinfo{person}{Baining Guo}, {and}
  \bibinfo{person}{Heung-Yeung Shum}.} \bibinfo{year}{2004}\natexlab{}.
\newblock \showarticletitle{Mesh editing with poisson-based gradient field
  manipulation}.
\newblock In \bibinfo{booktitle}{\emph{ACM SIGGRAPH 2004 Papers}}.
  \bibinfo{pages}{644--651}.
\newblock


\bibitem[Zheng et~al\mbox{.}(2021)]%
        {Zheng:secondary_motion:2021}
\bibfield{author}{\bibinfo{person}{Mianlun Zheng}, \bibinfo{person}{Yi Zhou},
  \bibinfo{person}{Duygu Ceylan}, {and} \bibinfo{person}{Jernej Barbic}.}
  \bibinfo{year}{2021}\natexlab{}.
\newblock \showarticletitle{A Deep Emulator for Secondary Motion of 3D
  Characters}. In \bibinfo{booktitle}{\emph{Proceedings of the IEEE/CVF
  Conference on Computer Vision and Pattern Recognition}}.
  \bibinfo{pages}{5932--5940}.
\newblock


\bibitem[Zhou and Jacobson(2016)]%
        {zhou2016thingi10k}
\bibfield{author}{\bibinfo{person}{Qingnan Zhou} {and} \bibinfo{person}{Alec
  Jacobson}.} \bibinfo{year}{2016}\natexlab{}.
\newblock \showarticletitle{Thingi10k: A dataset of 10,000 3d-printing models}.
\newblock \bibinfo{journal}{\emph{arXiv preprint arXiv:1605.04797}}
  (\bibinfo{year}{2016}).
\newblock


\bibitem[Zuffi et~al\mbox{.}(2017)]%
        {SMAL:2017}
\bibfield{author}{\bibinfo{person}{Silvia Zuffi}, \bibinfo{person}{Angjoo
  Kanazawa}, \bibinfo{person}{David Jacobs}, {and} \bibinfo{person}{Michael~J.
  Black}.} \bibinfo{year}{2017}\natexlab{}.
\newblock \showarticletitle{{3D} Menagerie: Modeling the {3D} Shape and Pose of
  Animals}. In \bibinfo{booktitle}{\emph{IEEE Conf. on Computer Vision and
  Pattern Recognition (CVPR)}}.
\newblock


\end{thebibliography}
\appendix
\pagebreak
\section{Invariance to the choice of frames}
\label{a:lemma}
\begin{claim}
Our framework is completely invariant to the (arbitrary) choice of frames $\set{\fram_i}.$
\end{claim}
This statement follows from two observations:
\begin{enumerate}
    \item The basis-dependent, projected jacobians play a role in only  two parts of our framework, namely, in the Poisson solve, Eq. \eqref{eq:lsq_poisson}, and in the loss $\jloss$, Eq. \eqref{eq:jloss}. In both cases they are used in exactly one type of function, of the form $\norm{\rest_i - Q_i}_F$, where  $Q_i = \nabla_{i} \gtmap$  is a ground-truth jacobian of a mapping $\gtmap$, expressed in the same basis $\fram_i$ as $\rest_i$, and $\norm{\cdot}_F$ is the frobenius norm.
    \item The choice of $\fram_i$ will not change the value of the expression $\norm{\rest_i - Q_i}_F$. This follows from basic linear algebra which we prove in the  lemma below.
\end{enumerate} 
Hence the only expression in which the frames appear is invariant to their choice, and therefore our method is completely invariant to the choice of frames.
\begin{unlemma}
Let $X,Y$ be linear transformations acting on a tangent space $\tangent$, $X,Y:\tangent\to\Real^3$, and let $\fram_1,\fram_2$ be two different orthogonal bases for $\tangent$. Let $X_1,Y_1\in\Real^{3\times2}$ (resp. $X_2,Y_2$) be the representation of  the linear transformations in the local coordinates of $\fram_1$ (resp. $\fram_2$). Then $\norm{X_1-Y_1}_F = \norm{X_2-Y_2}_F$.
\end{unlemma}
\begin{proof}
\begin{align*} 
\norm{X_1 - Y_1}_F \stackrel{FI}{=} \norm{\parr{X_1-Y_1}\fram_1^T\fram_2 }_F =\\
\norm{X_1\fram_1^T\fram_2-Y_1\fram_1^T\fram_2}_F \stackrel{CB}{=} \norm{X_2-Y_2}_F, 
\end{align*}
 where ``FI'' is due to the Frobenius norm's invariance to multiplication by orthogonal matrices ($\fram_1^T\fram_2$ is a $2\times2$ orthogonal matrix), and ``CB'' follows from the definition of a change of basis for a linear transformation.
\end{proof}

\clearpage

\end{document}